\documentclass[reprint,amsmath,amssymb,aps,prl,superscriptaddress, showpacs, eprint]{revtex4-2}
\usepackage[utf8]{inputenc}

\usepackage{amsmath,amssymb,ascmac,fancybox}
\usepackage{color}
\usepackage{graphicx}
\usepackage{url}
\usepackage{siunitx}
\usepackage{bm}
\usepackage{physics}
\usepackage{mathtools}
\usepackage{hyperref}
\usepackage[T1]{fontenc}
\usepackage{ulem}
\graphicspath{figures/}

\renewcommand{\vec}{\vb*}

\newcommand{\chris}[2]{\Gamma^{#1}_{#2}}

\newcommand{\lspin}{\lambda_s}

\newcommand{\lp}{l_C}

\begin{document}
\date{\today}
\title{Emergent curved space and gravitational lensing in quantum materials}
\author{Yugo Onishi}
\email{yugo0o24@mit.edu}
\affiliation{Department of Physics, Massachusetts Institute of Technology, Cambridge, MA 02139, USA}
\author{Nisarga Paul}
\email{npaul@mit.edu}
\affiliation{Department of Physics, Massachusetts Institute of Technology, Cambridge, MA 02139, USA}
\author{Liang Fu}
\affiliation{Department of Physics, Massachusetts Institute of Technology, Cambridge, MA 02139, USA}

\begin{abstract}
    We show that an effective gravitational field naturally emerges in quantum materials with long-wavelength spin (or pseudospin) textures. When the itinerant electrons' spin strongly couples to the background spin texture, it effectively behaves as a spinless particle in a curved space, with the curvature arising from quantum corrections to the electron's spin orientation. 
    The emergent curved space gives rise to the electron lensing effect, an analog of the gravitational lensing. The lensing effect can appear in systems without (emergent) magnetic fields, such as those with coplanar spin textures.
    Our work shows that novel ``gravitational'' phenomena generically appear in quantum systems due to nonadiabaticity, opening new research directions in quantum physics.   
\end{abstract}

\maketitle

\textit{Introduction. ---} A fascinating aspect of quantum physics is the emergence of new particles and fields in systems of electrons.  %
One example is the emergent gauge field. In a system of itinerant electrons strongly coupled to localized spins, a gauge field naturally emerges from the coupling between the electrons and the localized spins in the strong coupling limit, realizing an effective magnetic field acting on the itinerant electrons \cite{sundaram_wave-packet_1999, volovik_linear_1987, loss_berrys_1990, ye_berry_1999, taguchi_spin_2001, onoda_anomalous_2004}. This effective magnetic field further gives rise to the Hall effect without external magnetic fields \cite{neubauer_topological_2009, ohgushi_spin_2000, bruno_topological_2004, hamamoto_quantized_2015, addison_anomalous_2024, park_quantum_2025} and flat Chern bands that mimic Landau levels \cite{paul_giant_2023}.   %

\begin{figure}
    \centering
    \includegraphics[width=1.0\linewidth]{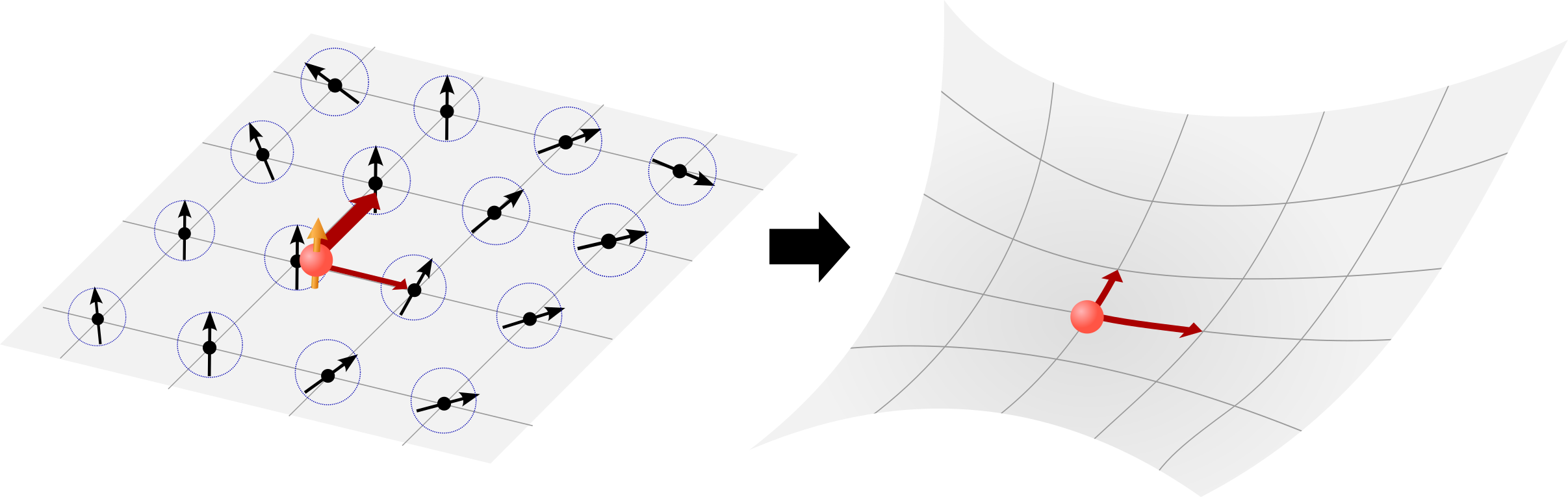}
    \caption{Schematic illustration of emergent curved space: a spinful electron in a slowly varying spin texture is equivalent to a spinless electron in a curved space, whose metric encodes the ``stretching'' of real-space distances by the spin texture.
    }
    \label{fig:spin_curved_space}
\end{figure}

In this work, we show that an effective gravitational field naturally emerges in quantum materials with long-wavelength spin (or pseudospin) textures by considering the correction from the strong coupling limit. Specifically, when electrons are coupled to localized spins, they are effectively described as spinless particles in a curved space, as depicted in Fig.~\ref{fig:spin_curved_space}. The curved space gives rise to an electron lensing effect, an analog of gravitational lensing~\cite{schneider_gravitational_1992}. This effect becomes more significant for fast electrons than slow electrons, in contrast to the scattering due to the effects of a scalar potential.

\textit{General theory. ---}
We consider itinerant electrons coupled with localized spins in general $d$-dimensions:
\begin{align}
    H = \frac{(\vec{p}-e\vec{A})^2}{2m} - J\vec{S}(\vec{x})\vdot\vec{\sigma}, \label{eq:original_Hamiltonian}
\end{align}
where $e(<0), m, \vec{p}, \vec{\sigma}$ are the charge, mass, momentum, and spin of the itinerant electrons, $\vec{A}$ is the vector potential describing the external magnetic field, 
and $\vec{S}(\vec{x})$ is the localized spin at position $\vec{x}$, with normalization $\abs{\vec{S}}=1$. $J$ is the coupling constant between the electrons and the localized spins. 
When $J$ is the largest energy scale, the spin of the itinerant electrons is effectively frozen so that their spins almost always align with the local spin $\vec{S}(\vec{x})$ and can be treated adiabatically.

When $J$ is finite,  high-order terms become important. Our focus is on the low-energy effective theory up to $\order{J^{-1}}$.
To %
derive the effective theory, we apply the unitary transformation to rotate electron's spin to align with the local axis:  
    $U^{\dagger}(\vec{x}) (\vec{S}(\vec{x})\vdot\vec{\sigma}) U(\vec{x}) = \sigma_z$. 
The new Hamiltonian $H'$ in the basis defined by $U(\vec{x})$ is~\cite{fujita_gauge_2011, tan_yangmills_2020}
\begin{align}
    H' &= U^{\dagger}(\vec{x}) H U(\vec{x}) = \frac{(\vec{p}-e\vec{A}-\hbar\vec{a})^2}{2m} - J\sigma_z. \label{eq:afterUnitary}
\end{align}
Here, $\vec{a} = U^{\dagger}i\nabla U$ is the $U(2)$ gauge field associated with $U(\vec{x})$. 
From Hamiltonian~\eqref{eq:afterUnitary}, we show that the effective description reduces to spinless electrons in curved space.

Before presenting the general theory, we consider a simple example of a spiral spin texture with wavevector $\vec{q}$: $\vec{S}(\vec{x}) = (\cos\vec{q}\vdot\vec{x}, \sin\vec{q}\vdot\vec{x}, 0)$.
Note that there is no emergent magnetic field since the spin texture is coplanar. 
In this case, we can choose the unitary transformation $U(\vec{x})$ as $U(\vec{x}) = \sigma_z\cos(\vec{q}\vdot\vec{x}/2) + \sigma_x\sin(\vec{q}\vdot\vec{x}/2)$ to diagonalize the spin-dependent term in Eq.~\eqref{eq:original_Hamiltonian}, and the resulting $U(2)$ gauge field is constant given by $\vec{a}=U^{\dagger}i\nabla U=-\vec{q}\sigma_y/2$.
We note that $\vec{a}$ \textit{cannot} be gauged away even though it is constant, because it is a $U(2)$ gauge field, not $U(1)$. Since $H'$ in this case is translationally invariant, it is easily diagonalized by the plane waves, and the resulting energy dispersion of the low-energy branch is
\begin{align}
    E(\vec{p}) &= \frac{p^2}{2m} + \frac{\hbar^2\Tr G}{2m} - \sqrt{J^2 + \frac{\hbar^2}{m^2}G_{ij} p_i p_j} \nonumber \\
    &= \frac{p_i g^{ij} p_j}{2m} + \frac{\hbar^2\Tr G}{2m} - J + \order{J^{-2}}, \label{eq:energy_spiral}
\end{align}
where the repeated indices are summed over the spatial indices. Here, $G_{ij} = a^{i}_{12}a^j_{21}=q_iq_j/4$
and we have defined the \textit{effective metric} $g^{ij}=\delta^{ij} - \hbar^2G_{ij}/(mJ)$.

The effective metric $g^{ij}$ in Eq.~\eqref{eq:energy_spiral} shows that the spin spiral suppresses the kinetic energy along $\vec{q}$ at order $\order{J^{-1}}$. This can be intuitively understood as the suppression of the hopping~\cite{de_gennes_effects_1960}: for large $J$, the itinerant electron's spin almost always aligns with the localized spins, hence the transfer integral is reduced along the varying spin directions, as indicated in Fig.~\ref{fig:spin_curved_space}. Correspondingly, the group velocity $\dv*{E}{\vec{p}}$ along $\vec{q}$ is reduced.

The reduction of the velocity offers a natural interpretation of the effective metric $g^{ij}$: we interpret it as the effective length of the real space becoming longer than it actually is. 
The corresponding real space metric is given by the inverse matrix of $g^{ij}$, which we denote by $g_{ij}$ and is given by $g_{ij}=\delta_{ij}+\hbar^2 G_{ij}/(mJ)$ up to $\order{J^{-1}}$.
$g_{ij}$ shows that the effective length $\dd{l}\equiv \sqrt{g_{ij}\dd{x}^i\dd{x}^j}$ along $\vec{q}$ becomes longer by a factor of $1+\hbar^2q^2/(4mJ)$.

We emphasize that the correction to the effective metric $g_{ij}$ appears only at $\order{J^{-1}}$. This should be contrasted with the conventional treatments~\cite{fujita_gauge_2011, tan_yangmills_2020, ohnishi_adiabatic_2007, addison_anomalous_2024, aharonov_origin_1992, berry_classical_1997, bruno_topological_2004}, which include an effective magnetic field or scalar potential up to $\order{J^0}$ but miss the effect in $g_{ij}$.

For general spin textures, we expect $g_{ij}$ to vary in space, giving rise to an effectively curved space. To derive this, we write
\begin{subequations}
\begin{align}
    H' &= \begin{pmatrix}
        H_1 & \Lambda \\
        \Lambda^\dagger & H_2
    \end{pmatrix},\\
    H_{i} &= \mp J + (\vec \pi_i^2 + |\vec a_{12}|^2)/2m,\\
    \Lambda &= -\{\vec{p}-e\vec{A},\vec a_{12}\}/2m,
\end{align}
\end{subequations}
with $\vec \pi_i = \vec p-e \vec A-\vec a_{ii}$, and we treat off-diagonal terms using a Schrieffer–Wolff (SW) transformation--- a unitary transformation decoupling the low and high-energy subspaces. At this point, it is useful to define three length scales: (1) $l_C = \hbar/\sqrt{mJ}$, the ``Compton wavelength'' associated to $J$ (we will see that $J\sim mc^2$ in a relativistic analogy), (2) $\lambda_s = |\nabla \vec S|^{-1}$, the scale over which $\vec S(\vec x)$ varies, and (3) $\lambda_F\sim\hbar/\sqrt{mE}$, the wavelength of the electron at energy $E$. The SW transformation yields the following low-energy leading-order effective Hamiltonian:
\begin{subequations}
\begin{align}
\mathcal{H} &= H_1 - \frac{\Lambda\Lambda^\dagger}{2J}\\
 &= H_1 - \frac{\acomm{\vec{p}-e\vec{A}}{\vec{a}_{12}}\acomm{\vec{a}_{21}}{\vec{p}-e\vec{A}}}{8m^2 J}, \label{eq:Heff_general}
\end{align}
\end{subequations}
which holds for {\it arbitrary} spin textures in the regime 
\begin{equation}
    l_C^2\ll \lambda_s\lambda_F,
\end{equation}
or equivalently $\langle\Lambda\rangle\ll J$ (details in Supplemental Materials (SM)). \par 

Furthermore, when $\vec{S}(\vec{x})$ varies sufficiently slowly such that $\lambda_s\gg \lambda_F$, the effective Hamiltonian simplifies, since derivatives of $\vec{a}_{12}, \vec{a}_{21}$ are negligible compared to the other terms. In this regime, which we adopt henceforth, the effective Hamiltonian up to $\order{l_C^2/\lspin^{2}}$ is given by
\begin{align}
    \mathcal{H} &= -J + \frac{\pi_{i}g^{ij}\pi_{j}}{2m} + V(\vec{x}), \label{eq:Heff}
\end{align}
where $g, V(\vec{x})$ are the effective metric and potential: 
\begin{subequations}
\begin{align}
    g^{ij} &= \delta^{ij} - \lp^2 G_{ij},  \label{eq:metric_inv_metric}\\  %
    V(\vec{x}) &= \frac{\hbar^2\Tr G}{2m}  - \frac{e\hbar\lp^2}{4m}\vec{\Omega}\vdot\vec{B}. \label{eq:Veff} 
\end{align}
\end{subequations}
Here, $G$ and $\vec\Omega$ are the quantum metric and the Berry curvature of the real space quantum geometry, respectively. They are defined with respect to the electron's spin state aligned with the local spin at position $\vec{R}$: $(\vec{S}(\vec{R})\vdot\vec{\sigma}) \ket{\psi_{\vec{R}}} = \ket{\psi_{\vec{R}}}$. By regarding $\vec{R}$ as a parameter, we define the quantum geometric tensor $Q_{ij}$ as 
\begin{align}
    Q_{ij}&=\ip{\partial_{R_i} \psi_{\vec{R}}}{\partial_{R_j}\psi_{\vec{R}}}-\ip{\partial_{R_i}\psi_{\vec{R}}}{\psi_{\vec{R}}}\ip{\psi_{\vec{R}}}{\partial_{R_j}\psi_{\vec{R}}}
\end{align}
We can readily show $Q_{ij}=a^{i}_{12}a^{j}_{21}$, and $G$ and $\vec{\Omega}$ are given by 
$G_{ij}=\Re Q_{ij}$ and $\Omega^i=-\epsilon^{ijk} \Im Q_{jk}$. $G$ can also be written with $S(\vec{x})$ as $G_{ij} \equiv \frac{1}{4}(\partial_i\vec{S})\vdot(\partial_j \vec{S})$.
We can verify that the Hamiltonian~\eqref{eq:Heff} reduces to Eq.~\eqref{eq:energy_spiral} for a spin spiral with uniform metric $g$.

Real space quantum geometry naturally enters the effective Hamiltonian~\eqref{eq:Heff} in two ways. The first is through the effective potential $V(\vec{x})$, whose leading $\mathcal{O}(l_C^0)$ term is proportional to $\Tr G\propto\abs{\nabla\vec{S}}^2$~\cite{bruno_topological_2004} and corresponds to the ``geometric potential'' in literature~\cite{berry_classical_1997, aharonov_origin_1992, ohnishi_adiabatic_2007}. The subleading $\mathcal{O}(l_C^2)$ term in $V(\vec{x})$ represents orbital magnetization coupling to the external $\vec B$ field. 

Interestingly, quantum geometry also manifests in a second way: through the effective metric $g^{ij}$. The quantum metric $G_{ij}$ directly modifies the spatial metric $g^{ij}$ and thereby influences electrons' motion. The inverse of $g^{ij}$, denoted by $g_{ij}$, defines the effective line element $\dd{l}=\sqrt{g_{ij}\dd{x}^i\dd{x}^j}$, and thus a spatially varying quantum metric leads to an effective curved space, i.e., \textit{emergent gravitational field}~\footnote{We note that the Hamiltonian for a particle in curved space can commonly take a different form, but we can show that the two formulations are equivalent up to $\order{l_C^2\lspin^{-4}}$. See SM for more details.}. 
The effect of the emergent curved space scales as $\mathcal{O}(l_C^2/\lambda_F^2)\sim \mathcal{O}(E_K/J)$, in close analogy with true (post-Newtonian/Einstein) gravitational effects scaling as $\mathcal{O}(E_K/mc^2)$, where $E_K$ is the kinetic energy of a particle of mass $m$ and $c$ is the speed of light. Hence $J\sim mc^2$ in this analogy, which we make precise using a Lagrangian formulation in SM.

Importantly, the emergent curved space associated with the spatial metric $g$ gives rise to physical effects that cannot be captured by a scalar potential. This parallels how Einstein’s theory of gravity goes beyond Newtonian gravity. We now discuss physical consequences of the emergent gravitational field that are qualitatively new.

\textit{Gravitational lensing. ---}
To understand the physical effects of $g(\vec{x})$, we analyze the classical equations of motion of the electrons, given by Hamilton's equations: 
\begin{subequations}
    \begin{align}
    \dot{x}^i &= \pdv{\mathcal{H}}{p_i} = \frac{g^{ij}\pi_j}{m}, \label{eq:xdot} \\
    \dot{p}_i &= -\pdv{\mathcal{H}}{x^i} = -\frac{1}{2m}\pi_j \pdv{g^{jk}}{x^i}\pi_k + \frac{e}{m}\pdv{\mathcal{A}_j}{x^i}g^{jk}\pi_k- \pdv{V}{x^i}, \label{eq:pdot} 
\end{align}
\end{subequations}
where $\vec{\mathcal{A}}=\vec{A}+(\hbar/e)\vec{a}_{11}$ is the total effective vector potential, including the external and emergent gauge fields. 
Taking the time derivative of Eq.~\eqref{eq:xdot} and using Eq.~\eqref{eq:pdot}, we find 
\begin{align}
    m(\ddot{x}^i + \Gamma^{i}_{jk}\dot{x}^j\dot{x}^k) = F^i, \label{eq:EOM_general}
\end{align}
where $\chris{i}{jk}=(g^{il}/2)(\partial_jg_{kl}+\partial_kg_{jl}-\partial_{l}g_{jk})$ is the Christoffel symbol with $g_{ij}$ the inverse of $g^{ij}$. $F^i=g^{ij}(e\mathcal{B}_{jk}\dot{x}^k - \partial_j{V})$ is the force from the potential $V$ and the total magnetic field $\mathcal{B}_{jk}=\partial_j \mathcal{A}_k -\partial_k \mathcal{A}_j$. The term with $\chris{i}{jk}$ scales as $\mathcal{O}(l_C^2/\lambda_F^2)$ and captures the effects of the emergent curved space. Notably, the electron feels a force proportional to the square of its velocity.\par

For constant effective potential $V$ and zero magnetic field $\mathcal{B}$, the right hand side of Eq.~\eqref{eq:EOM_general} vanishes, yielding the geodesic equation: $\ddot{x}^i + \Gamma^{i}_{kl}\dot{x}^k \dot{x}^l=0$. The electron thus follows a geodesic in the curved space in this case.
For constant metrics, $\Gamma^i_{jk}$ vanishes and the solutions are straight lines, while for spatially varying metrics, $\Gamma^{i}_{jk}\neq 0$ and thus the geodesics are no longer straight lines in general, i.e., the emergent curved space bends the trajectory. Notably, such trajectories are independent of the particle's initial velocity up to a rescaling of time: 
if $\vec{x}(t)$ is a geodesic, so is $\vec{x}(\alpha t)$ for an arbitrary constant $\alpha$.
Therefore, the effects of $g$ persist even for fast-moving particles. This is in sharp contrast with the force from a potential $V$, 
which is independent of velocity 
and thus affects fast-moving particles less than slow particles. 
This is analogous to how the trajectory of light
is affected by a curved space-time metric (Einstein gravity) but not by a scalar potential (Newtonian gravity).

\par 
Let us discuss perhaps the simplest consequence of the emergent curved space: the bending of electron trajectories between domains of distinct spin orders. For instance, suppose $\vec S(\vec x) = \hat z$ in the region $x\to -\infty$ and 
    $\vec S(\vec x) = (0, \sin \frac{2\pi x}{\lspin}, \cos \frac{2\pi x}{\lspin})$
in the region $x\to +\infty$, in two spatial dimensions. This describes the transition between ferromagnetic and helical domains, which could be realized for instance in a material with the Dzyaloshinskii-Moriya interaction. The emergent metric is trivially $g^{ij} = \delta^{ij}$ in the ferromagnetic region, while $g^{xx}$ receives a correction
    $g^{xx} = 1 -(\pi l_C/\lspin)^2$,
in the helical region. The resulting electron motion can be worked out using the conservation of energy $E = m g_{ij}\dot x^i \dot x^j/2$  and conservation of $y$-momentum $p_y = mg_{yi}\dot x^i$. We ignore the potential $V$ for now. This can be justified, for instance, by electrostatic gating in the helical region. Then, letting $\theta_{L/R}$ denote the angles of a geodesic with respect to the $x$-axis in the left and right regions, respectively, one finds 
\begin{equation}
    \frac{\tan\theta_R}{\tan\theta_L} = \sqrt{1+ (\pi l_C/\lspin)^2}.
\end{equation}
This describes the bending of electrons towards the normal as they pass from the ferromagnetic region to the helical region. In contrast with Snell's law $\sin\theta_R/\sin\theta_L = n_L/n_R$ for light, there is no condition for total internal reflection. This effect remains even in the presence of the potential $V$, and admits interesting generalizations, for instance to the case of domain walls between distinct spiral phases, 
which we discuss in SM. 

\par 
Next, as a simple and solvable example, we consider the spin texture which we call the \textit{radial spiral}
\begin{equation}
    \vec S(\vec x) = (\cos \frac{2\pi r}{\lambda_s},\sin \frac{2\pi r}{\lambda_s}, 0),
    \label{eq:spin_cone}
\end{equation}
where $r =|\vec x|$ is the radial coordinate. Locally, the radial spiral approximates configurations minimizing the spin gradient energy $(\partial_i\vec S)^2$ under radial boundary conditions $\vec S(r_{\text{min}})\neq \vec S(r_{\text{max}})$. Globally, the radial spiral is the unique continuous, rotationally symmetric, coplanar spin texture with constant $V(\vec x)$ (see SM for more details).

\begin{figure}
    \centering    \includegraphics[width=0.8\linewidth]{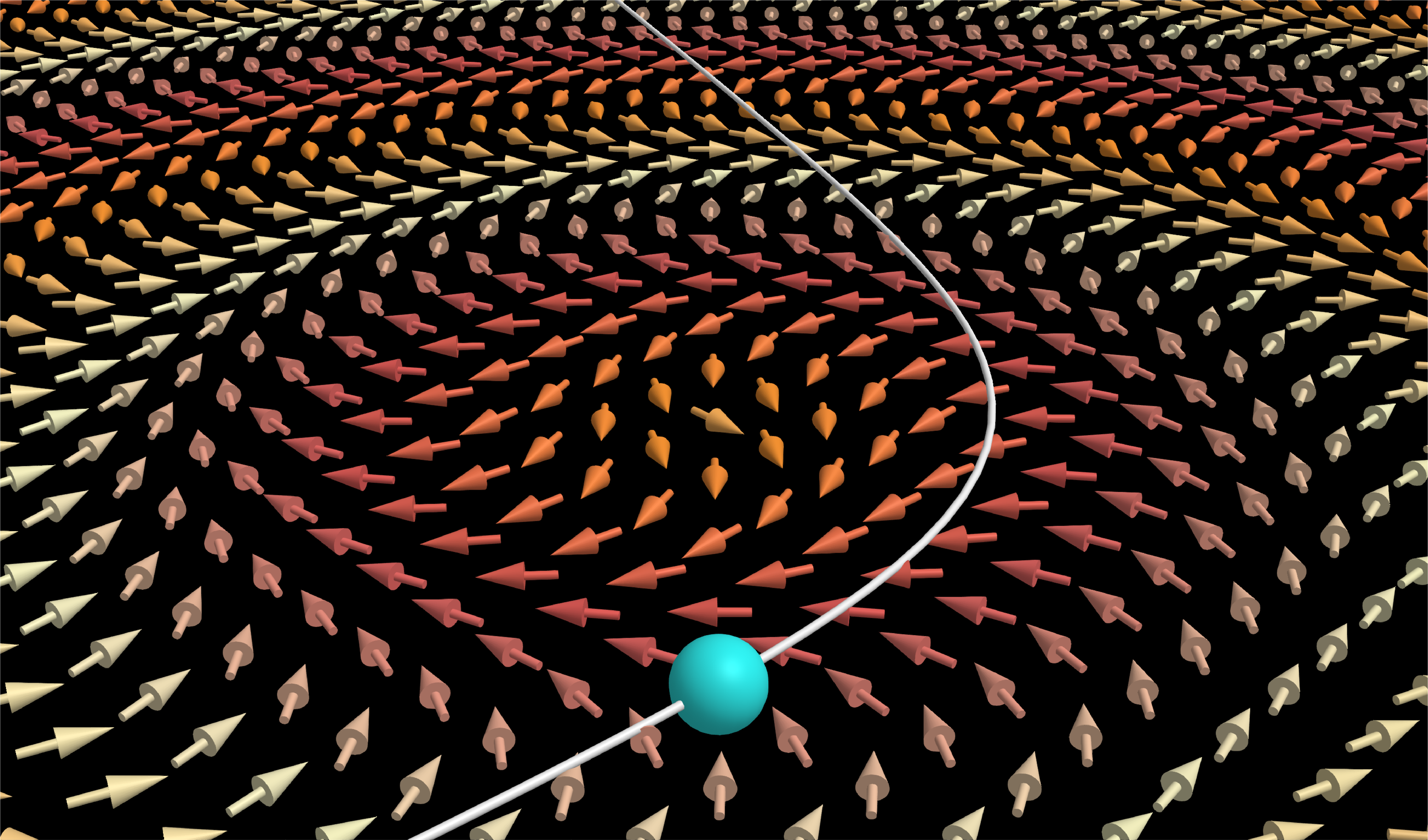}
    \caption{The radial spiral spin texture (Eq.~\eqref{eq:spin_cone}) and a corresponding electron geodesic (solid curve). The effectively curved space gives rise to gravitational lensing of electrons around the origin.}    \label{fig:conical}
\end{figure}

The radial spiral is depicted in Fig.~\ref{fig:conical}. It satisfies $\vec b =0$, $V(\vec x) = h^2/8m\lspin^2$, and
\begin{equation}\label{eq:g_cone}
    g^{ij} = \delta^{ij} -\left(\frac{\pi l_C}{\lambda_s}\right)^2\frac{ x^ix^j}{r^2}.
\end{equation}
Electron motion is thus described simply by the geodesic equation, which takes the form
\begin{equation}\label{eq:cone_geodesic}
    \ddot{\vec x} = -\left(\frac{\pi l_C}{\lspin}\right)^2 \frac{(\dot{\vec x}\times \vec x)^2\vec x}{x^4},
\end{equation}
indicating an $\mathcal{O}(l_C^2/\lspin^2)$ correction to free-particle motion. 
The geodesics are $r(t)=\pm \sqrt{(v_0t+r_0)^2+(\xi ht/r_0)^2}$, $\theta(t)=\theta_0 + (\tan^{-1}(\gamma t + \delta) - \tan^{-1} \delta)/\xi$ in polar coordinates with $\xi = \sqrt{1-(\pi l_C/\lspin)^2}$, where $h = r^2\dot\theta$ is a conserved angular momentum, $(r(0),\dot r(0),\theta(0)) = (r_0,v_0,\theta_0)$, $\gamma = (h\xi/r_0^2)(1+\delta^2)$ and $\delta = r_0v_0/(h\xi)$. 
Alternatively, the geodesics can be expressed as straight lines in coordinates which trivialize the metric, as shown in SM.  
An example trajectory is plotted in Fig.~\ref{fig:conical}, showing clearly the \textit{lensing} of the electron geodesic. \par 

The properties of geodesics in the previous two examples can be elucidated by a geometric perspective. Quite generally, the emergent curved space in $d$ spatial dimensions corresponds to that of a surface embedded in higher dimensions. This follows from the effective line element $\dd{l}=\sqrt{g_{ij}\dd{x}^i\dd{x}^j}$, which can be rewritten as
\begin{align}
    \dd{l}^2 = \dd{\vec{x}}^2 + \lp^2 G_{ij}\dd{x}^i\dd{x}^j = \dd{\vec{x}}^2 + \lp^2\dd{s}^2. \label{eq:line_elements}
\end{align}
Here, $\dd{s}=\sqrt{G_{ij}\dd{x}^i \dd{x}^j}$ defines the quantum distance via the quantum metric $G_{ij}$ in the local Hilbert space, which is $\mathbb{C}P^1\simeq S^2$ in our case. 
The effective line element Eq.~\eqref{eq:line_elements} thus corresponds to a $d$-dimensional %
surface embedded in the ($d+2$)-dimensional manifold, $\mathbb{R}^d\times\mathbb{C}P^1$, with $\lp$ the length scale of the extra dimensions. 
For coplanar textures such as  $\vec S(\vec x) = (\cos\theta(\vec x),\sin\theta(\vec x),0)$, the electron's spin is restricted to a one-dimensional subspace, and the effective curved space is a surface in $d+1$ dimensions. The effective line element is given by
\begin{equation}\label{eq:g_coplanar}
    \dd{l}^2 = \dd{\vec{x}}^2+\frac{\lp^2}{4}\dd{\theta}^2 = \dd{\vec{x}}^2+\dd{z}_{\text{eff}}^2,
\end{equation}
whose geometry is the $d$-dimensional surface defined by $(x^1,\dots,x^d, z_{\text{eff}}(\vec{x}))$ with $z_{\text{eff}}=\lp\theta(\vec x)/2$. 

\par 
From this geometric perspective, the metric and geodesics of the ferromagnet-to-helical transition discussed above correspond to the geometry of two half-planes joined at $x=0$, with $z_{\text{eff}} = 0$ and $z_{\text{eff}} = \pi l_Cx/\lspin$ in the left and right regions, respectively. For the radial spiral, the metric and the geodesics correspond to the geometry of a cone with $z_{\rm eff}=\pi\lp r/\lambda_s$.
Incidentally, the cone geometry Eq.~\eqref{eq:g_cone} is the solution of the Einstein equations with a point mass $M$ in (2+1)-dimensions with the identification $(\pi l_C/\lambda_s)^2 \sim 8GM/c^2$ where $G$ is the gravitational constant~\cite{Deser1984Jan}. The radial spiral therefore realizes the simplest non-vacuum solution of (2+1)d Einstein gravity with vanishing cosmological constant. More general spin textures generically yield more complex metrics and geodesics. We collectively refer to the ensuing new effects on electronic motion as \textit{gravitational lensing}. Crucially, gravitational lensing becomes stronger for faster electrons and cannot be explained by an effective magnetic or electric field.

\begin{figure}
    \centering
    \includegraphics[width=0.7\linewidth]{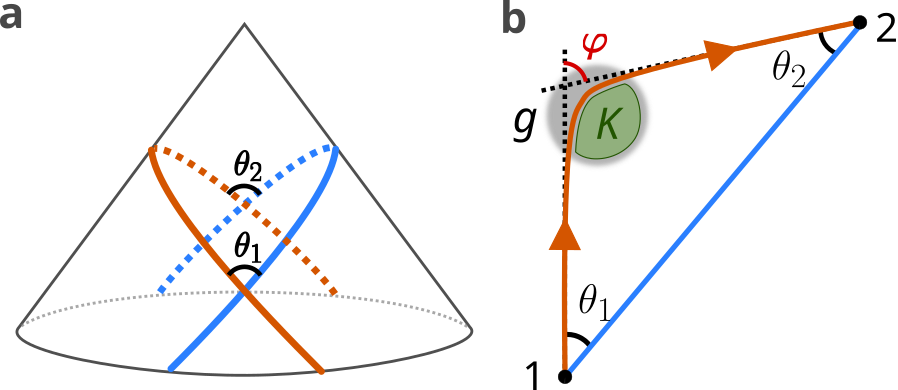}
    \caption{(a) Two geodesics on a cone. (b) Deflection due to gravitational lensing when the curved metric and the curvature are localized in a gray region. The deflection angle $\varphi$ is given by the curvature $K$ enclosed by the blue and orange lines.}
    \label{fig:deflection}
\end{figure}

Curvature of the effective curved space is essential in gravitational lensing. In flat space with zero curvature, there exist coordinates where geodesics, i.e., trajectories of free-particle motion, are straight lines, and any two geodesics can intersect at most once. Thus, multiple intersections between geodesics require finite curvature in the effective space. 
\par
In two-dimensional systems, this observation is formalized with the Gauss-Bonnet theorem, which relates a closed loop on a curved surface to the Gaussian curvature enclosed by the loop~\cite{nakahara_geometry_2018, kobayashi_differential_2019}. 
For a polygon $S$ made of geodesics on a curved surface, the following relation follows from the Gauss-Bonnet theorem:
\begin{align}
    \sum_{i} \phi_i + K = \sum_{i} (\pi-\theta_i) + K = 2\pi, \label{eq:Gauss_Bonnet}
\end{align}
where $i$ is the label for the vertices, $\phi_i$ and $\theta_i=\pi-\phi_i$ are the exterior and interior angle at vertex $i$, and $K$ is the integrated Gaussian curvature enclosed by the polygon:
\begin{align}
    K &= \int_S\kappa\sqrt{\det g_{ij}}\dd^2x.
\end{align}
Here, $\kappa=R_{xyxy}/\det g_{ij}$ is the Gaussian curvature and $R_{ijkl}$ is the Riemann curvature tensor for the effective metric $g_{ij}$ 
(see SM). 
We note that $R_{ijkl}$ can be expressed in terms of the Riemann curvature of the Hilbert space defined by the quantum metric, as detailed in SM.
For the coplanar texture Eq.~\eqref{eq:g_coplanar}, this takes the form
\begin{equation}\label{eq:K_for_coplanar}
    K = \int\dd^2{x} \frac{\lp^2}{4} \frac{ (\partial_x^2\theta)(\partial_y^2\theta)- \qty(\partial_x\partial_y \theta)^2}{(1 + \abs{\lp\partial_i\theta/2}^2)^{3/2}}.
\end{equation}
When the metric is trivial and the curvature vanishes, $K=0$ and Eq.~\eqref{eq:Gauss_Bonnet} reduces to the well-known polygon exterior angle sum theorem. On a curved surface, $K$ is finite, and the exterior angle sum will be modified. 

Eq.~\eqref{eq:Gauss_Bonnet} applies when two geodesics intersect at two points as in Fig.~\ref{fig:deflection}(a), since they form a two-edged polygon. In this case, Eq.~\eqref{eq:Gauss_Bonnet} reduces to
\begin{align}
    \theta_1 + \theta_2 &= K. \label{eq:crossing}
\end{align}
It should be emphasized that $\theta_i$ is measured using the effective metric $g_{ij}$, and may differ from the usual angle defined by the trivial metric $\delta_{ij}$. For the radial spiral, $\theta_i$'s are measured on the corresponding cone, as shown in Fig.~\ref{fig:deflection}. %
The curvature vanishes except at the origin, which contributes to $K$ by the angle deficit of the cone: $\kappa = 2\pi(1-1/\sqrt{1+(\pi\lp/\lambda_s)^2})\delta(\vec{x})$.

More generally, the deflection angle is related to the curvature. Consider the case where the curved metric (and thus the curvature) is localized in a small region, as shown in Fig.~\ref{fig:deflection}(b), and a particle travels from point 1 to point 2 on a geodesic (orange line). If the points are sufficiently far away from the curved region, we can also connect the points only through the flat space with a geodesic, which is a straight line (blue line). Then we can define the deflection angle $\varphi$ as the angle change of the velocity from point 1 to 2 within the flat region. On the other hand, since the orange and blue lines form a geodesic polygon, we can apply Eq.~\eqref{eq:crossing}. Since the deflection angle is given by $\varphi=\theta_1+\theta_2$, it follows that $\varphi = K$.
Note that $\varphi$ here is the angle change of the velocity measured at points 1 and 2 within the flat space, and thus is defined with the flat metric $\delta_{ij}$.   
This relation reveals the geometric nature of the gravitational lensing~\footnote{After completing this work, we found similar relations for the true gravitational lensing of light in Ref.~\cite{gibbons_applications_2008}.}. 

For example, we may consider a coplanar texture varying with wavelength $\lambda_s$. From Eq.~\eqref{eq:K_for_coplanar}, the deflection angle over a region of area $A$ scales as
    $\varphi \sim Al_C^2 /\lambda_s^4$.
For masses ranging from $0.1$ to 1 electron mass and typical $10$-$100$~meV Hund's couplings, $l_C\sim 1$-$10$~nm, while typical spin textures have $\lspin \sim 10$-$100$~nm. Thus, the deflection angle from gravitational lensing is within the observable range over areas $A\sim \lambda_s^2$, and can be enhanced by coherent electron motion across larger samples.

\textit{Discussion. ---}
The emergent curved space we proposed originates from the quantum nature of electrons. Indeed, the spatial metric correction vanishes as $\hbar\to 0$. It can also be interpreted through the electron's wave nature. The effective metric $g$ determines the electron’s wavelength $\lambda$ for a given frequency $\omega$, and thus plays the role of the refractive index.
A spatially varying $g$ corresponds to a spatially dependent refractive index, and thus causes refraction, i.e., the lensing effect. We note that gravitational lensing of light in general relativity can also be understood with an effective refractive index~\cite{schneider_gravitational_1992}.

It is important to note that the emergent curved space --- and thus the gravitational lensing --- does not require any symmetry breaking. This contrasts with the emergent magnetic field, which requires the time reversal symmetry breaking. In particular, gravitational lensing can occur without magnetic fields in contrast to the conventional electron lensing in electron microscopes~\cite{egerton_physical_2016}, where magnetic fields are used to control electrons' trajectory. Gravitational lensing does not require noncoplanar spin textures either, as we demonstrated with the radial spiral. This is in sharp contrast to emergent magnetic field~\cite{sundaram_wave-packet_1999, volovik_linear_1987, loss_berrys_1990, ye_berry_1999, taguchi_spin_2001, onoda_anomalous_2004}.

A similar gravitational picture is expected to hold for general systems described by a Hamiltonian of the form~\eqref{eq:afterUnitary} with $SU(2)$ gauge fields, under assumptions analogous to $\lp^2 \ll \lambda_s\lambda_F$ and $\lambda_s\gg \lambda_F$. For instance, the spin degrees $\vec{\sigma}$ can be replaced with a pseudo-spin. We also note that our assumptions on the length scales allow only perturbatively small gravitational fields ($\abs{g_{ij}-\delta_{ij}}\ll 1$). Exploring analogues of strong gravitational effects in a similar setup, such as a black hole, would be an interesting future direction. 

A curved space was discussed in contexts of condensed matter previously. For example, strain responses or deformation effects~\cite{avron_viscosity_1995, hughes_torsional_2011, bradlyn_kubo_2012, dong_geometrodynamics_2018, rao_hall_2020, dahlhaus_geodesic_2010} and phonon scattering by a vortex~\cite{fischer_riemannian_2002} were discussed with a curved space picture. Weyl semimetals with inhomogeneous tilt of the Weyl cone, which may be induced by strain or external fields, were also studied in analogy with the curved spacetime~\cite{volovik_black_2016, weststrom_designer_2017, liang_curved_2019, huang_black-hole_2018, zubkov_black_2018, zubkov_emergent_2015, guan_artificial_2017, nissinen_type-iii_2017, liang_curved_2019} and gravitational lensing~\cite{konye_anisotropic_2023, haller_black_2023}. We emphasize, however, that the emergent curved space proposed here does \textit{not} require external strain or nontrivial band structure. It emerges due to the quantum mechanical coupling between itinerant electrons and localized spins, and would be potentially as diverse as the spin textures, offering rich platforms. The dynamics of spin textures~\cite{barnes_generalization_2007, nagaosa_topological_2013, yang_universal_2009, yamane_continuous_2011, yamane_electric_2016} may further lead to time-dependent curved space, such as a gravitational wave analog~\cite{chojnacki_gravitational_2024}, which we leave for future study.
A curved space is also used as a theoretical tool to study electronic systems, such as fractional quantum Hall systems~\cite{wen_shift_1992, avron_viscosity_1995, estienne_ideal_2023}. The emergent curved space 
may provide a method to experimentally test these theoretical results.

The concept of emergent gauge field is so far well established and generically appears in various quantum systems. In these systems, there should always be a correction due to the finite energy gap, which results in emergent gravitational fields.
Therefore, we expect emergent curved space to appear in quantum systems ubiquitously.

\acknowledgements 
YO thanks Filippo Gaggioli for encouraging discussion. LF and NP thank Aidan Reddy for participation in a related project. This work was supported by the Air Force Office of Scientific Research under award number FA2386-24-1-4043. LF was supported in part by the Simons Investigator Award from the Simons Foundation. 

\clearpage
\appendix
\section*{Supplemental Materials}
\section{Derivation of the effective Hamiltonian}\label{app:eff}
\subsection{Effective Hamiltonian for general spin textures}
Consider the following Hamiltonian: 
\begin{align}
    H &= \frac{(\vec{p}-e\vec{A})^2}{2m} - J\vec{S}(\vec{x})\vdot\vec{\sigma}
\end{align}
Here, we include the spatial dependence of $J$. 
In this section, assuming $J$ is large, we derive the low-energy effective theory up to $\order{J^{-1}}$
by applying the Schrieffer-Wolff transformation. 

We first consider the unitary matrix $U(\vec{x})$ that diagonalizes the spin:
\begin{align}
    U^{\dagger}(\vec{x}) (\vec{S}(\vec{x})\vdot\vec{\sigma}) U(\vec{x}) &= \sigma_z 
\end{align}
Then we can obtain Hamiltonian $H'$ in the new basis defined by $U(\vec{x})$:
\begin{align}
    H' &= U^{\dagger}(\vec{x}) H U(\vec{x}) = \frac{(\vec{p}-e\vec{A}-\vec{a})^2}{2m} - J\sigma_z
\end{align}
where we set $\hbar=1$, and $\vec{a}$ is defined as 
\begin{align}
    \vec{a} &= U^{\dagger}(\vec{x})i\nabla U(\vec{x}) = \begin{pmatrix}
        \vec{a}_{11} & \vec{a}_{12} \\
        \vec{a}_{21} & \vec{a}_{22}
    \end{pmatrix}
\end{align}
and serves as a $U(2)$ gauge field. From now on, we include $(\vec{a}_{11}+\vec{a}_{22})/2$ in the regular vector potential $\vec{A}$ and treat $\vec{a}$ as a $SU(2)$ gauge field with $\vec{a}_{11}+\vec{a}_{22}=0$. Noting that $\vec{a}_{12}=\vec{a}_{21}^*$, $H'$ can be written as 
\begin{align}
    H' 
    &= H_0 + V
\end{align}
where 
\begin{subequations}
\begin{align}
    H_0 &= \begin{pmatrix}
        H_1 & 0 \\
        0 & H_2 
    \end{pmatrix} \\
    V &= \begin{pmatrix}
        0 & \Lambda \\
        \Lambda^{\dagger} & 0
    \end{pmatrix} \\
    \Lambda 
    &= -\frac{1}{2m}\acomm{\vec{\pi}_0}{\vec{a}_{12}}
\end{align}
\end{subequations}
Here, $H_i = (\vec{\pi}_{i}^2 + \abs{\vec{a}_{12}}^2)/(2m) \mp J$, $\vec{\pi}_i = \vec{p}-e\vec{A}-\vec{a}_{ii}$, and $\vec{\pi}_0=\vec{p}-e\vec{A}$. It may be worth noting that $\Lambda$ has a similar form to the strain generator $J^{\mu}_{\ \nu}\equiv \acomm{x^{\mu}}{p_{\nu}}$~\cite{bradlyn_kubo_2012, rao_hall_2020}. When $\vec{a}_{12}$ is linear in the position $x^i$ and $\vec{A}=0$, then $\Lambda$ coincides with the strain generator up to a factor. For more general $\vec{a}_{12}$, $\Lambda$ can be interpreted as a generator for a position-dependent strain.

We now apply Schrieffer–Wolff transformation to $H'$ assuming that $J$ is large. We consider a unitary transformation $e^{S}$ to apply to $H'$ as $H''=e^{S}H'e^{-S}$. Assuming that $S=S^{(1)} + S^{(2)} + \dots$ with $S^{(i)}=\order{(\Lambda/J)^{i}}$, $H''$ is expanded as
\begin{align}
    H'' &= H' + \comm{S}{H'} + \frac{1}{2}\comm{S}{\comm{S}{H'}} + \dots \nonumber\\
    &= H_0 + V + \comm{S}{H_0} + \comm{S}{V} + \frac{1}{2}\comm{S}{\comm{S}{H_0}} + \dots
\end{align}
We require $S^{(1)}$ to satisfy the following:
\begin{align}
    V + \comm{S^{(1)}}{H_0} &= 0.
\end{align}
Writing each element explicitly, 
\begin{subequations}
\begin{align}
    S_{11}^{(1)}H_1 - H_{1}S_{11}^{(1)} &= 0, \\
    \Lambda + S_{12}^{(1)}H_2 - H_1 S_{12}^{(1)} &= 0, \\
    \Lambda^{\dagger} + S_{21}^{(1)}H_1 - H_2S_{21}^{(1)} &= 0, \\
    S_{22}^{(1)}H_2 - H_{2}S_{22}^{(1)} &= 0.
\end{align}
\end{subequations}
For the leading order in $\Lambda/J$, these can be solved as
\begin{align}
    S_{12}^{(1)} &= -S_{21}^{(1)\dagger} = -\frac{\Lambda}{2J}, \label{ap:eq:S12} \\
    S_{11}^{(1)} &= S_{22}^{(1)} = 0.
\end{align}
Then the effective Hamiltonian for the low energy sector up to $\order{J^{-1}}$ is given by $H_{\rm eff}=H''_{11}$ as 
\begin{align}
    H_{\rm eff} 
    &= - J + \frac{\vec{\pi}_{1}^2 + \abs{\vec{a}_{12}}^2}{2m}  - \frac{\Lambda\Lambda^{\dagger}}{2J} \nonumber \\
    &= - J + \frac{\vec{\pi}_{1}^2 + \abs{\vec{a}_{12}}^2}{2m}  - \frac{\acomm{\vec{\pi}_0}{\vec{a}_{12}}\acomm{\vec{a}_{21}}{\vec{\pi}_0}}{8m^2 J} \label{ap:eq:Heff1}
\end{align}
As we discussed in the main text, this effective Hamiltonian is valid when $\lp^2\ll\lambda_s\lambda_F$ with $\lp\equiv1/\sqrt{mJ}$, the length scale of the spin texture $\lspin\equiv \abs{\nabla \vec{S}}^{-1}$, and the Fermi wavelength $\lambda_F$.

\subsection{When the spin is slowly varying in space}
When the length scale of spatial variation of $\vec{S}(\vec{x})$, denoted as $\lspin$, is large so that $\lspin \gg \lambda_F$, the derivative of $\vec{a}_{12}$ in Eq.~\eqref{ap:eq:Heff1} is negligible compared to the other terms. Therefore, we can further simplify Eq.~\eqref{ap:eq:Heff1} as 
\begin{align}
    H_{\rm eff} &= - J + \frac{\vec{\pi}_{1}^2 + \abs{\vec{a}_{12}}^2}{2m}  - \frac{\pi_{0i}Q_{ij}\pi_{0j}}{2m^2 J} 
    + \order{E_F\frac{\lambda_F\lp^2}{\lspin^3}} \nonumber \\
    &= - J + \frac{\vec{\pi}_{1}^2 + \abs{\vec{a}_{12}}^2}{2m}  - \frac{\pi_{0i}G_{ij}\pi_{0j}}{2m^2 J} - \frac{e\hbar^3}{4m^2J}\vec{\Omega}\vdot\vec{B}
\end{align}
where $Q_{ij}=a_{12}^{i}a_{21}^{j}$ is the quantum geometric tensor, $G_{ij}=\Re Q_{ij}$, $\Omega_{ij}=-2\Im Q_{ij}$ are the quantum metric and the Berry curvature respectively, and $\vec{\Omega}$ is defined as $\Omega^{i}=\epsilon^{ijk}\Omega_{jk}/2$. $E_F$ is the Fermi energy. Since $\vec{a}_{11}=\order{1/\lspin}$ and $G_{ij}=\order{1/\lspin^2}$, we can replace $\vec{\pi}_0$ with $\vec{\pi}_1$ up to the same accuracy of the approximation to obtain 
\begin{align}
    H_{\rm eff} &= -J(\vec{x}) + \frac{\pi_{1i}g^{ij}\pi_{1j}}{2m} + \frac{\hbar^2\Tr G}{2m}  - \frac{e^3\hbar}{4m^2 J}\vec{\Omega}\vdot\vec{B} \nonumber \\
    &\quad + \order{E_F\frac{\lambda_F\lp^2}{\lspin^3}}
\end{align}
Here, $g^{ij}= \delta^{ij} - \lp^2G_{ij}$ is the effective metric, and $\vec{\Omega}=\curl\vec{a}_{11}$ is the Berry curvature.

\section{Analogy to true gravity}\label{app:lagrange}

The emergent gravity is an analogue of Einstein gravity in condensed matter. This analogy becomes clearer in the Lagrangian formalism. 
The Lagrangian corresponding to the effective Hamiltonian in the main text is given by $L=p_i\dot{x}^i-\mathcal{H}$ as 
\begin{align}
    L(x, \dot{x}) = \frac{m}{2}g_{ij}(\vec{x})\dot{x}^{i}\dot{x}^{j} + e\mathcal{A}_i(\vec{x}) \dot{x}^i + J - V(\vec{x}) \label{eq:Lagrangian}
 \end{align}
This Lagrangian can be obtained from the following action up to $\order{J^{-1}\lspin^{-2}}$:
\begin{align}
    S &= S_0 + m\tilde{c}\int\sqrt{\tilde{g}_{\mu\nu}dx^{\mu}dx^{\nu}} \label{eq:action}
\end{align}
Here, $S_0=\int\dd{t}(m\dot{x}^2/2 - V + e\mathcal{A}_i\dot{x}^i)$ is the Lagrangian for a particle in a flat space-time with potential $V$ and vector potential $\mathcal{A}_i$. 
The second term in Eq.~\eqref{eq:action} represents the effect of the emergent gravity, where $\mu$ and $\nu$ run over $0$ to $3$, $(x^0, x^1, x^2, x^3) = (\tilde{c}t, x, y, z)$. $\tilde{c}$ is defined through $m\tilde{c}^2 = J$. The space-time metric $\tilde{g}$ is given by 
\begin{align}
    \tilde{g}_{\mu\nu} = \begin{pmatrix}
        1 &  \\
        & \frac{\hbar^2}{m^2\tilde{c}^2}G_{\mu\nu}
    \end{pmatrix}
\end{align}

Eq.~\eqref{eq:action} shows that the correction to our low-energy effective action takes the same form as the action for a particle with the space-time metric $\tilde{g}$. The role of the rest energy $mc^2$ in the true gravity with $c$ the speed of light is played by $J=m\tilde{c}^2$ in the emergent gravity, defining an effective speed of light $\tilde{c}$. 
The effect from the spatial components of the space time metric gives the correction to the Hamiltonian of order $\order{E_F/(m\tilde{c}^2)}$, just as the true gravity effects from the curved space-time scales as $\order{E_K/(mc^2)}$ with $E_K$ the kinetic energy of a particle.

\section{Correspondence to other forms of Hamiltonian in a curved space}\label{app:correspondence}
In the main text, we discuss the effective Hamiltonian in the following form as the Hamiltonian in a curved space: 
\begin{align}
    H^{(1)} &= \frac{p_i g^{ij} p_j}{2m} + V(\vec{x}) \label{ap:eq:H(1)}
\end{align}
In literature, there is another form of the Hamiltonian which also describes an electron in curved space given by 
\begin{align}
    H^{(2)} &= \frac{1}{2m} \frac{1}{\sqrt{g}} p_i (\sqrt{g}g^{ij} p_j) + V(\vec{x}) \label{ap:eq:H(2)}
\end{align}
where $g = \det g_{ij}$. The latter is obtained by replacing the Laplacian in the Hamiltonian in flat space with the one in a curved space. In this section, we show that these are equivalent up to the accuracy we discussed in the main text. 

The most important difference between Eq.~\eqref{ap:eq:H(1)} and \eqref{ap:eq:H(2)} is the normalization of the wavefunction $\psi(\vec{x})$ they assume. In the main text, we always assume that the wavefunction is normalized in the following way:
\begin{align}
    \int\dd^d x \abs{\psi^{(1)}(\vec{x})}^2 = 1
\end{align}
On the other hand, the Hamiltonian $H^{(2)}$ implicitly assumes that the wavefunction is normalized as follows:
\begin{align}
    \int\dd^dx \sqrt{g} \abs{\psi^{(2)}(\vec{x})}^2 = 1 
\end{align}
These two conventions are related as follows: the solution of the Schrodinger equation under $H^{(1)}$, $\psi^{(1)}$, is related to the one under $H^{(2)}$, $\psi^{(2)}$, as $\psi^{(1)}=g^{1/4}\psi^{(2)}$. Since $g$ depends on position in general, the states $\psi^{(1)}$ and $\psi^{(2)}$ and their corresponding Hamiltonian look different from each another. 

To show their equivalence, we formulate the Schrodinger equation in a variational way. The Schrodinger equation under Hamiltonian $H^{(i)}$ is obtained from the following action:
\begin{align}
    S^{(1)}[\psi] &= \int\dd{t}\dd^d{x} \qty[i\psi^*\pdv{\psi}{t}-\frac{1}{2m}(p_{i}\psi)^* g^{ij} (p_j\psi)-V\abs{\psi}^2] \\
    S^{(2)}[\psi] &= \int\dd{t}\dd^d{x}\sqrt{g} \qty[i\psi^*\pdv{\psi}{t}-\frac{1}{2m}(p_{i}\psi)^* g^{ij} (p_j\psi)-V\abs{\psi}^2] 
\end{align}
where we have set $\hbar=1$. We find the Schrodinger equation under $H^{(i)}$ by requiring $S^{(i)}$ to be stationary under arbitrary variations of $\psi$ and $\psi^{*}$: $\delta S^{(i)}/\delta\psi^{*}=0$, and its complex conjugate. 

Now we would like to show that $S^{(1)}[g^{1/4}\psi]=S^{(2)}[\psi]$ up to the accuracy we considered in the main text. We start with $S^{(1)}[g^{1/4}\psi]$ and rewrite it as follows:
\begin{align}
    S^{(1)}[g^{1/4}\psi] = &\int\dd{t}\dd^d{x} \sqrt{g}\left[i\psi^*\pdv{\psi}{t} -V\abs{\psi}^2 \right. \nonumber \\
    &\left. -\frac{1}{2m}(g^{-1/4}p_{i}g^{1/4}\psi)^* g^{ij} (g^{-1/4}p_jg^{1/4}\psi)\right]
\end{align}
Here, we have assumed $g$ is time-independent. Noting $g^{-1/4}p_i g^{1/4}=p_i - i\alpha_i$ with $\alpha_i = (1/4)\partial_i \log g$, we can rewrite $S^{(1)}[g^{1/4}\psi]$ as 
\begin{align}
    S^{(1)}[g^{1/4}\psi] = &\int\dd{t}\dd^d{x} \sqrt{g}\left[i\psi^*\pdv{\psi}{t} -V\abs{\psi}^2\right. \nonumber \\
    &\left. -\frac{1}{2m}((p_{i}-i\alpha_i)\psi)^* g^{ij} ((p_j-i\alpha_j)\psi)\right] \nonumber \\
    &= S^{(2)}[\psi] + \int\dd{t}\dd^d{x} \sqrt{g}\frac{1}{2m} (\partial_i\alpha^i-\alpha^i\alpha_i)\abs{\psi}^2 \label{ap:eq:S1_S2}
\end{align}
where $\alpha^i = g^{ij}\alpha_j$. Eq.~\eqref{ap:eq:S1_S2} shows that the theory~\eqref{ap:eq:H(1)} differs from the theory~\eqref{ap:eq:H(2)} only by a correction to a potential, $\delta V=(\partial_i\alpha^i-\alpha^i\alpha_i)/(2m)$. 

Lastly, we note that the metric $g$ we considered in the main text is $g_{ij} = \delta_{ij} + \order{\lp^2\lspin^{-2}}$ with $\lp=(mJ)^{-1/2}$ the coupling constant and $\lspin$ the length scale of the spin texture. Therefore, $\alpha=\order{\lp^2\lspin^{-3}}$ and thus the last term in Eq.~\eqref{ap:eq:S1_S2} is $\order{\lp^2\lspin^{-4}}$. Then we find
\begin{align}
    S^{(1)}[g^{1/4}\psi] = S^{(2)}[\psi] + \order{\lp^2\lspin^{-4}}.
\end{align}
Therefore, $H{(1)}$ and $H^{(2)}$ are equivalent up to $\order{\lp^2\lspin^{-4}}$.

\section{Geodesics in spin textures}\label{app:spin}
In this section, we provide more details on the various spin textures explored in the main text and extensions thereof. 
\subsection{Radial spiral spin texture}

Here, we present details on the radial spiral spin texture
\begin{equation}
    \vec S(\vec x) = (\cos \frac{2\pi }{\lspin}r,\sin \frac{2\pi }{\lspin}r,0),
\end{equation}
where $r=\sqrt{x^2+y^2}$, a simple spin texture with constant real-space quantum metric and $\vec b=0$. This combination of features implies that only the emergent gravitational field
\begin{equation}
    g^{ij} = \delta^{ij} -\nu\frac{ x^ix^j}{r^2}.
\end{equation}
enters the electron equation of motion. Here we've defined
\begin{equation}
    \nu = (\pi l_C/\lspin)^2.
\end{equation}

\subsubsection{Geodesics from direct calculation}
The geodesic equation $\ddot x^i + \Gamma^{i}_{jk} \dot x^j \dot x^k = 0$
reduces to 
\begin{equation}
\begin{aligned}
    \ddot{\vec x} &= -\nu \vec x \frac{(\dot{\vec x} \times \vec x)^2}{x^4}.
\end{aligned}
\end{equation}
In radial coordinates, this is 
\begin{equation}
        \ddot{\vec r}= -\nu r\dot\theta^2 \hat{\vec r}.
\end{equation}
Comparing with 
\begin{equation}
    \ddot{\vec r} = (\ddot r - r\dot\theta^2)\hat{\vec r} + \dv{t}(r^2\dot\theta) \hat{\vec \theta},
\end{equation}
we find two scalar equations: 
\begin{subequations}
\begin{align}
    0 &= \dv{t} (r^2\dot\theta)\\
    \ddot r &= (1-\nu)r\dot\theta^2.
\end{align}
\end{subequations}
We may define a conserved angular momentum $h=r^2\dot\theta=$ constant. The remaining equation is 
\begin{equation}
    \ddot r = (1-\nu) h^2/r^3.
\end{equation}
Assuming $r(0) = r_0$ and $\dot r(0) = v_0$, we have
\begin{equation} r(t) = \pm \sqrt{\left(
   v_0t + r_0\right){}^2+(1-\nu)(ht/r_0)^2}.
\end{equation}
We can integrate to find $\theta(t)$ using $\dot \theta = h/r^2$. Let $\theta(0) = \theta_0$. 
We find 
\begin{equation}
\begin{aligned}
    &\theta(t) -\theta_0 = \\
   & \frac{1}{\sqrt{1-\nu}} \left(\tan^{-1}\left(\left[\frac{v_0^2}{h\sqrt{1-\nu}} + \frac{h\sqrt{1-\nu}}{r_0^2}\right]t+\frac{r_0v_0}{h\sqrt{1-\nu}}\right)\right. \\
   &\qquad \qquad \qquad -\left.\tan^{-1}\left(\frac{r_0v_0}{h\sqrt{1-\nu}}\right)\right)
\end{aligned}
\end{equation}
This describes a particle on a cone, with the conical singularity at $r=0$. This geometry also describes the solution of the Einstein equations in the presence of a point mass in (2+1)d with zero cosmological constant~\cite{Deser1984Jan}. Note that there is no Newtonian attraction; a geodesic at fixed $r(t) = r_0$ is perfectly valid. However, there is ``focusing"; two trajectories which pass on either side of the origin will intersect on the other side. 
\par
\begin{figure}
    \centering    \includegraphics[width=0.8\linewidth]{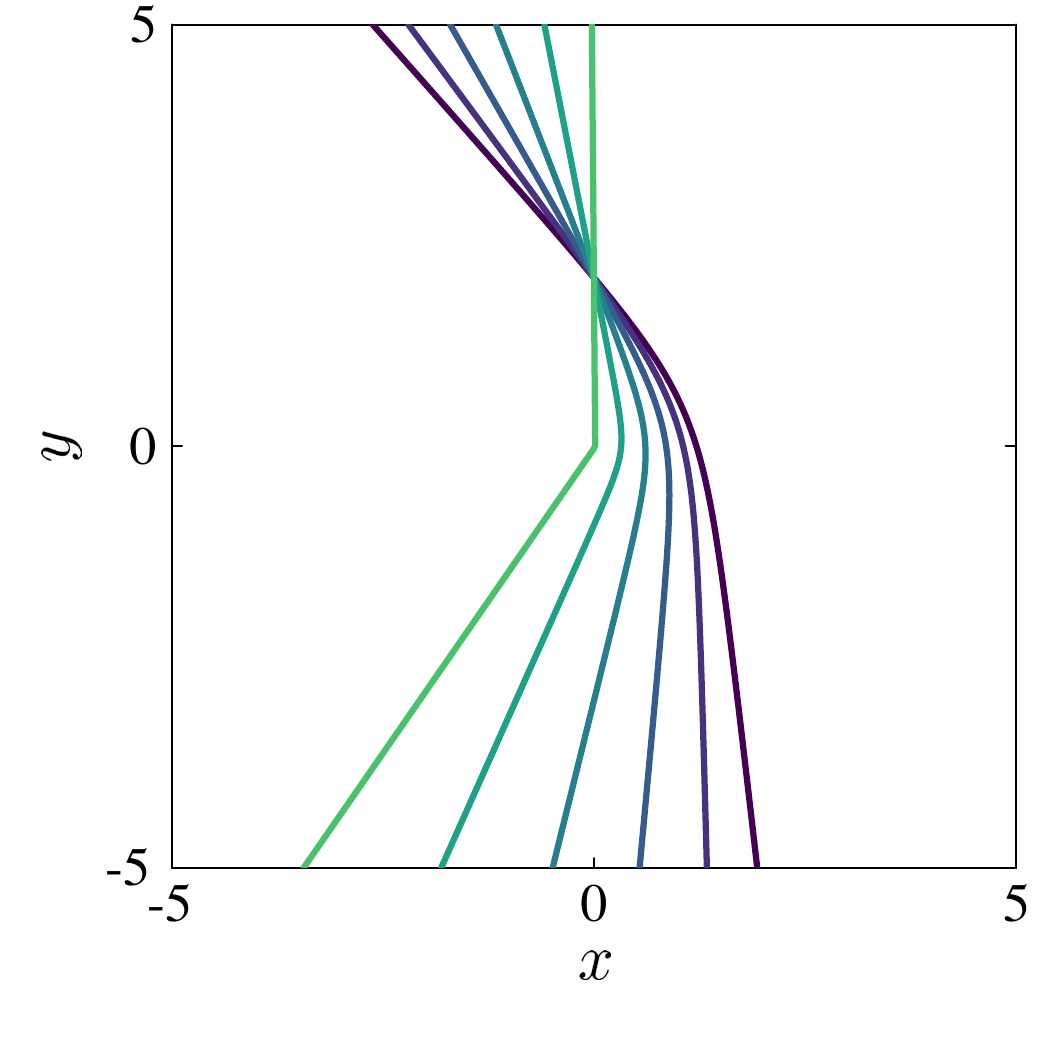}
    \caption{Example geodesics  of the cone geometry with $\nu = 0.3$ (corresponding to a deficit angle $\approx 0.24\pi$) from Eq.~\eqref{eq:geodesics_v2}. These satisfy $r_0 = 2, \theta_0 = \pi/2$, and varying $\phi_0$. As the geodesics approach the origin, they become the union of two rays.}
    \label{fig:geodesics}
\end{figure}

 
At this point, geodesics depend on the independent parameters $(r_0,\theta_0,v_0,h)$. Another useful form is in terms of $(r_0,\theta_0,v_{\text{i}},\phi_0)$ where 
\begin{equation}
    \vec v_\text{i} = v_{\text{i}}(\cos\phi_0,\sin\phi_0)
\end{equation}
is the initial Cartesian velocity. The conserved angular momentum is
\begin{equation}
    h = r_0 v\sin(\phi_0-\theta_0),
\end{equation}
while the initial radial velocity $v_0$ satisfies
\begin{equation}
v_0^2 = v_{\text{i}}^2\cos^2(\phi_0-\theta_0). 
\end{equation}
We rescale time $t\mapsto t/v_{\text{i}}$ in the following, which omits explicit dependence on velocity. Defining the angle difference
\begin{equation}
    \vartheta_0 = \phi_0 -\theta_0,
\end{equation}
the geodesics are
\begin{equation}\label{eq:geodesics_v2}
    \begin{aligned}
        r(t) &= \pm \sqrt{(r_0 + t\cos \vartheta_0)^2 +(1-\nu) t^2 \sin^2(2\vartheta_0)/4}\\
        \theta(t) - \theta_0 &= \frac{1}{\sqrt{1-\nu}}\left(\tan^{-1}( \gamma t + \delta) -\tan^{-1}\delta\right)
    \end{aligned}
\end{equation}
where 
\begin{equation}
\begin{aligned} 
    \gamma &= -\frac{\cot(\vartheta_0)}{2\sqrt{1-\nu} r_0} ((1-\nu)\cos(2\vartheta_0)+\nu-3)\\
    \delta &= \frac{\csc\vartheta_0}{\sqrt{1-\nu}}.
\end{aligned}
\end{equation}
Example geodesics are plotted in Fig.~\ref{fig:geodesics} for $\nu = 0.3$. 

\subsubsection{Geodesics from trivial coordinates} 
\begin{figure}
    \centering
    \includegraphics[width=1.0\linewidth]{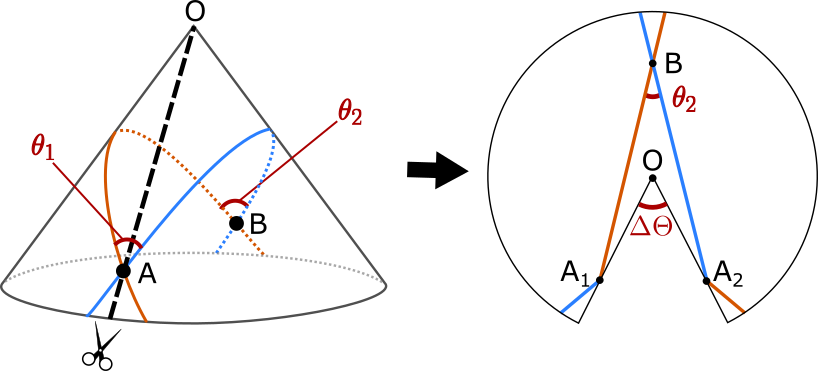}
    \caption{Geodesics on a cone and the corresponding net. The left panel shows geodesics on the cone intersecting at points A and B, with interior angles $\theta_1$ and $\theta_2$. The right panel shows the net obtained by cutting the cone along line OA, where points A$_1$ and A$_2$ both represent the same point A on the cone. The angle $\Delta\Theta$ denotes the cone’s angle deficit.}
    \label{ap:fig:cone_net}
\end{figure}

Alternatively, we can solve the geodesic equation for the radial spiral spin texture by taking coordinates in which the metric is trivial. To find such coordinates, we start with the effective line element for the radial spiral:
\begin{align}
    \dd{l}^2 &= dx^2 + dy^2 + \dd{z}_{\text{eff}}^2,
\end{align}
where $z_{\text{eff}}$ is the embedding coordinate. For the radial spiral, $z_{\text{eff}}=\frac{\pi l_C}{\lspin}r$ where $r=\sqrt{x^2+y^2}$.  
Therefore, in the polar coordinates $(r, \theta)$ with $r\ge 0, 0\le \theta < 2\pi$, the line element is given by 
\begin{align}
    \dd{l}^2 &= \qty(\frac{\dd{r}}{1-\alpha})^2 + r^2\dd{\theta}^2,
\end{align}
where $\alpha = 1-1/\sqrt{1+(\pi l_C/\lspin)^2}$. Then we can see that the metric is trivial in a coordinate $(R, \Theta)=(r/(1-\alpha), (1-\alpha)\theta)$. In this new coordinate, the line element becomes 
\begin{align}
    \dd{l}^2 &= \dd{R}^2 + R^2 \dd{\Theta} = \dd{X}^2 + \dd{Y}^2,
\end{align}
where $X=R\cos\Theta, Y=R\sin\Theta$. Therefore, in these new coordinates $(R,\Theta)$ or $(X, Y)$, the metric is that of a flat space. Note that $\Theta$ now only takes values between $0\le \Theta < 2\pi(1-\alpha)$ with angle deficit  $\Delta\Theta \equiv 2\pi \alpha$. Intuitively, the effective curved space of a radial spiral is equivalent to a cone, and $(X,Y)$ is the coordinate on the cone's net. Correspondingly, the deficit of $\Theta$ from $2\pi$, $\Delta\Theta$, is the angle deficit of the cone (also see Fig.~\ref{ap:fig:cone_net}). 
\par 
Since the metric in $(X, Y)$ coordinates is trivial, the trajectory of a particle is a straight line with constant velocity in $(X, Y)$ coordinates: $\vec{X}(t)=\vec{X}_0 + \vec{V}t$. Such straight lines can be expressed with $R, \Theta$ as 
\begin{align}
    R\cos(\Theta-\Theta_0) = \mathrm{const}.
\end{align}
with a constant $\Theta_0$. Rewriting this with $r, \theta$ coordinates, we find 
\begin{align}
    r\cos((1-\alpha)(\theta-\theta_0)) = \mathrm{const}\label{ap:eq:geo_radial_spiral}.
\end{align}

When a particle comes from $r=\infty$ at $t=-\infty$, passes by around the origin and goes to $r=\infty$  at $t=+\infty$, $\theta$ at $t=\pm \infty$ has to satisfy either $(1-\alpha)(\theta-\theta_0)=\pm\pi/2$. Therefore, $\theta$ at $t=\pm\infty$ is either $\theta_{\pm} = \pm\frac{\pi}{2(1-\alpha)} + \theta_0$. In particular, the deflection angle $\varphi$ is given by:
\begin{align}
    \abs{\varphi} = \pi - (\theta_{+}-\theta_{-}) = \frac{\pi\alpha}{1-\alpha}.
\end{align}
This deflection angle is clearly seen in Fig.~\ref{fig:geodesics}. \par 

We can also confirm the relation between the interior angles $\theta_1, \theta_2$ and the Gauss curvature $K$ in the main text, $\theta_1+\theta_2=K$, for the radial spiral by considering the geodesics on the cone's net. Consider when two geodesics intersect at two points A and B with interior angles $\theta_1$ and $\theta_2$, as illustrated in the left panel of Fig.~\ref{ap:fig:cone_net}. We can obtain the corresponding net by cutting the cone along line OA, which is shown in the right panel. Noting that the sum of the interior angles of a polygon OA$_2$BA$_1$ is $2\pi$ and that $\angle$OA$_1$B$+$$\angle$OA$_2$B$=\theta_1$, we find $\Delta\Theta = \theta_1 + \theta_2$. Since the integrated Gaussian curvature $K$ enclosed by the two geodesics is nothing but the angle deficit $\Delta\Theta$, we confirm $\theta_1 + \theta_2 = K$. 
We also note that the relation between the deflection angle $\varphi$ and the Gauss curvature $K$ in the main text, $\varphi=K$, does not apply to the current case, because the curved metric in the radial spiral is not localized as we assumed in deriving the relation.

While the deflection of geodesics in a curved space is qualitatively different than scattering off of a potential, we may note the following: deflection of geodesics in the cone geometry, and indeed any geometry with positive Gaussian curvature, resembles scattering due to a potential well, while deflection due to a negative curvature would resemble scattering off of a potential barrier. This follows from the Gauss-Bonnet theorem: when curvature is positive, a pair of geodesics can close to form a polygon, while when the curvature is negative this is not possible. When electrons scatter due to a localized spin texture (outside of which spins are uniform), the potential $V$ always acts as a barrier, while the emergent metric may scatter electrons in a manner resembling either a potential barrier or well (with Fig.~\ref{fig:geodesics} depicting the latter).

\subsubsection{Other properties of the radial spiral spin texture}
Let us discuss two properties of the radial spiral spin texture: (1) uniqueness properties and (2) approximation of realistic spin textures obtained from minimizing spin gradient energy. \par 

\textit{Uniqueness. } The radial spiral is the unique spin spiral which is coplanar, continuous, rotationally symmetric, and has constant squared gradient $\sum_i (\partial_i \vec S)^2$. In fact, this is true for the analog of the radial spiral in higher than two dimensions as well, though here we prove it for $d=2$. Uniqueness here is implicitly up to a spatially constant rotation in spin space, $\vec S(\vec x) \mapsto R \vec S(\vec x)$ where $R \in SO(3)$. \par 
To prove this, first let us apply a rotation so that $\vec S \cdot \hat z = 0$. Then we can assume
\begin{equation}
    \vec S(\vec x) = (\cos f(r),\sin f(r),0)
\end{equation}
in order to maintain $|\vec S| =1$ and rotational symmetry. Then $f''(r)=$ constant, which implies $f(r) =ar+b$, and we may assume $b=0$ by applying an appropriate rotation.\par

\textit{Minimizing spin gradient energy. } In the main text, we mentioned that the radial spiral approximately describes configurations minimizing the spin gradient energy in geometries with radial boundary conditions $\vec S(r_{\text{min}})\neq \vec S(r_{\text{max}})$. Here, we explain this further. \par 

Consider the problem of finding the normalized spin texture $\vec S(\vec x)$ which minimizes the spin gradient energy 
\begin{equation}
    E[\vec S] = \frac{K}{2} \int_\Omega \,[\dd \vec x]  (\partial_i \vec S)^2
\end{equation}
in a disk $\Omega$ with radius $r_{\text{max}}$ subject to the following boundary conditions: $\vec S(r=r_{\text{max}}) = \vec S_1$ and $\vec S(r\leq r_{\text{min}}) = \vec S_0$ for some $r_{\text{min}}< r_{\text{max}}$. For instance, this could describe a region of a ferromagnet outside of which spins point up, but inside of which spins are forced to point in-plane due to a local Zeeman field. \par 
Minimum energy configurations wind only in the $\vec S_1,\vec S_0$ plane. Without loss of generality, we assume $\vec S_1 = \hat y, \vec S_0 = \hat x$, and take (for two dimensions)
\begin{equation}\label{eq:Sxfrtheta}
    \vec S(\vec x) = (\cos f(r,\theta),\sin f(r,\theta),0).
\end{equation}
The energy density is $(\partial_r f)^2 + r^{-2} (\partial_\theta f)^2$. Since $\theta$-dependence only increases energy, we take $f(r,\theta) = f(r)$ and find $E[\vec S] = \frac{K}{2} \int_{r_{\text{min}}}^{r_{\text{max}}} \, \dd r\, r (\partial_r f)^2$. From the Euler-Lagrange equation $\partial_r (r \partial_r f) = 0$ it follows that $f(r) = a \ln r + b$, and matching boundary conditions yields 
\begin{equation}
    f(r) = \frac{\pi}{2} \frac{\ln(r/r_{\text{min}})}{\ln(r_{\text{max}}/r_{\text{min}})}. 
\end{equation}
At any radius $r_{\text{min}} < r_* < r_{\text{max}}$, $f(r)$ can of course be locally approximated to linear order as 
\begin{equation}
    f(r) \approx \frac{\pi}{2} \frac{\ln(r_*/r_{\text{min}})}{\ln(r_{\text{max}}/r_{\text{min}})} + \frac{\pi}{2r_*\ln(r_{\text{max}}/r_{\text{min}})} (r-r_*)
\end{equation}
and agrees with the radial spiral. The analogous energy minimization problem in $d$ dimensions yields $f(r) \sim r^{2-d}$, and a linear approximation once again is valid locally, agreeing with the radial spiral. In this sense, the radial spiral \textit{locally} describes an energy minimizing configuration with appropriate radial boundary conditions.

\subsection{Ferromagnet/ spin-helical domain wall} 

\begin{figure}
    \centering
\includegraphics[width=0.45\linewidth]{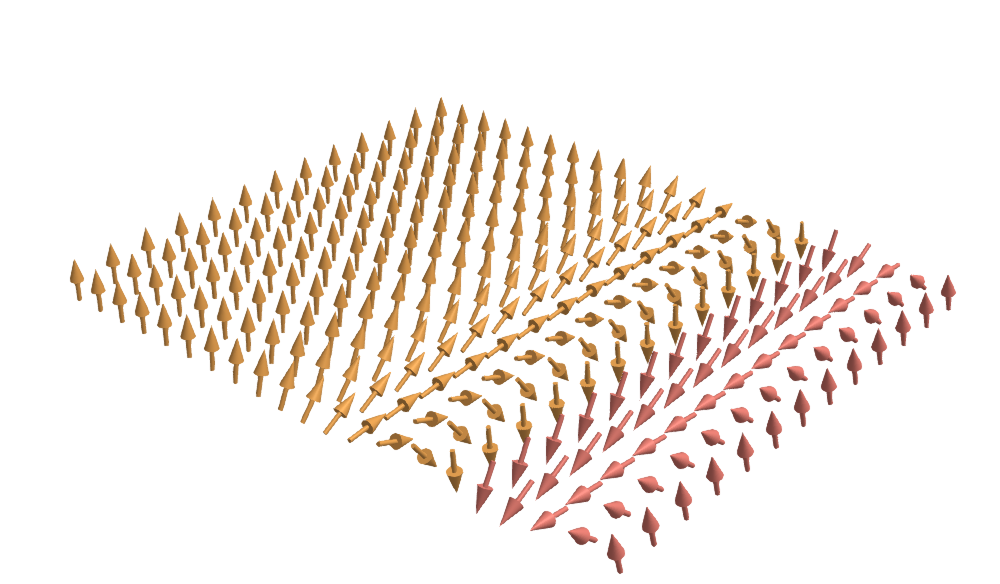}
\includegraphics[width=0.5\linewidth]{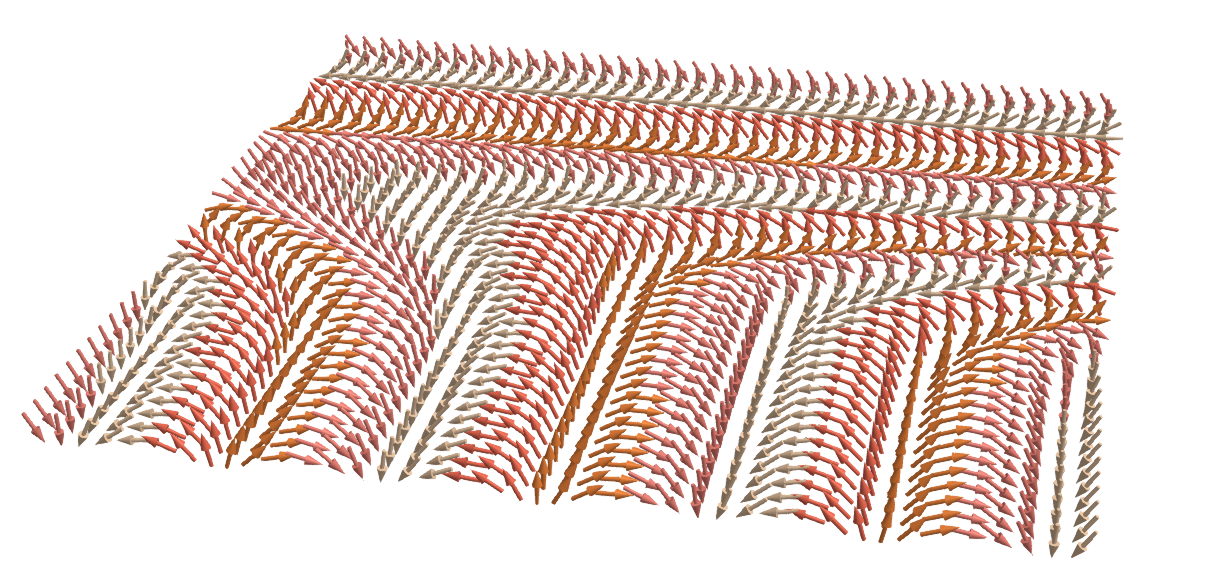}
    \caption{Left: the spin texture of Eq.~\eqref{eq:ferro_spiral}. Right: the spin texture of Eq.~\eqref{eq:spiral_spiral}.}
    \label{fig:ferro_spiral}
\end{figure}

We expand upon the discussion of the transition between ferromagnet and spin-helical regions in the main text. This is a simple setup which captures the bending of geodesics. We take
\begin{equation}\label{eq:ferro_spiral}
    \begin{aligned}
        \vec S(\vec x) &= (0,\sin f(\vec x), \cos f(\vec x)) \\
        f(\vec x) &= f(x) = \begin{cases}
            0 & x \to -\infty \\
            \frac{2\pi x}{\lspin} & x\to +\infty 
        \end{cases}. 
    \end{aligned}
\end{equation}
We plot this in Fig.~\ref{fig:ferro_spiral}. This sort of texture could arise in a material with a natural tendency to a helical phase (e.g. with the DMI) and a ferromagnet phase (strong Zeeman field) near a first-order transition; or in the helical phase with an inhomogenous Zeeman field which turns one side of the material into a ferromagnet. \par

It is straightforward to see that geodesics will bend across the boundary $x=0$. The metric is 
\begin{equation}
    g^{ij} = \delta^{ij} - \frac{l_C^2}{4} f'(x)^2\delta^{ix}\delta^{jx}.
\end{equation}
Geodesics conserve the energy
\begin{equation}
    E = \frac12 mg_{ij} \dot x^i \dot x^j. 
\end{equation}
To see this, note that 
\begin{equation}
    \begin{aligned}
        \dot E/m &= \frac12 \dot x^k \partial_k g_{ij} \dot x^i \dot x^j + g_{ij} \ddot x^i \dot x^j \\
        &= \frac12 \dot x^k (\Gamma^m_{ki}g_{mj} + \Gamma^m_{kj} g_{mi}) \dot x^i \dot x^j + g_{ij} \ddot x^i \dot x^j\\
        &= (g_{ij} \ddot x^i + \Gamma^{m}_{ki}g_{mj}\dot x^k \dot x^i) \dot x^j\\
        &= \dot x^j g_{rj} (\ddot x^r + \Gamma^r_{ki} \dot x^k \dot x^i) \\
        &= 0. 
    \end{aligned}
\end{equation}
Hence
\begin{equation}
    E = \frac12m \left(1+\frac14 l_C^2f'(x)^2\right)\dot x^2 + \frac12m\dot y^2 = \text{const.}
\end{equation}
Because the system has a $y$ translation invariance, $p_y = mg_{yi}\dot x^i$ is conserved. Hence $\dot y =$ constant and also
\begin{equation}
   \left(1+\frac14 l_C^2f'(x)^2\right)\dot x^2 = \text{const.}
\end{equation}
Let $v_{L/R}$ denote the $x$-velocity for $x\to \mp \infty$. It follows that 
\begin{equation}
    \left( 1+ \frac{\pi^2l_C^2}{\lspin^2}\right)v_R^2 = v_L^2
\end{equation}
giving us the relation
\begin{equation}
    \frac{v_R}{v_L} = \frac{1}{\sqrt{1+ (\pi l_C/\lspin)^2}}.
\end{equation}
Let $\theta_{L/R}$ be the angle of the geodesic with respect to the $x$ axis in the left and right regions, respectively. It follows that 
\begin{equation}
\frac{\tan\theta_R}{\tan\theta_L} = \sqrt{1+ (\pi l_C/\lspin)^2}.
\end{equation}
While this agrees with Snell's law $\sin\theta_L/\sin\theta_R = n_R/n_L$ for light traveling between two isotropic media with indices of refraction $n_{L/R}$ for \textit{small} angles $\theta_{L/R}$, it differs for larger angles. This distinction has the important consequence that there is no condition for total internal reflection for the relation above.\par 
So far, we've been ignoring the effect of $V(\vec x) = \hbar^2 f'(x)^2/2m$. If we include the effect of $V$, the geodesics instead conserve the energy 
\begin{equation}
    E = \frac12 m g_{ij}\dot x^i \dot x^j + V
\end{equation}
and following the same steps, 
\begin{equation}
    \frac12 \left(1+(\pi l_C/\lspin)^2\right)v_R^2 + \frac{h^2}{2m^2\lspin^2} = \frac12 v_L^2. 
\end{equation}
The resulting relation between asymptotic angles is modified to
\begin{equation}
    \frac{v_R}{v_L} = \frac{\tan\theta_L}{\tan\theta_R} = \frac{\sqrt{1-\frac{h^2}{m^2\lspin^2v_L^2}}}{\sqrt{1+\frac{\pi^2l_C^2}{\lspin^2}}}.
\end{equation}
Hence both $V$ and $g^{ij}$ contribute to the bending of trajectories. We can isolate the effect from $g^{ij}$ by considering a domain wall between two distinct spiral phases, which we turn to next. 

\subsection{Spiral / spiral domain wall}

\begin{figure}
    \centering
\includegraphics[width=0.8\linewidth]{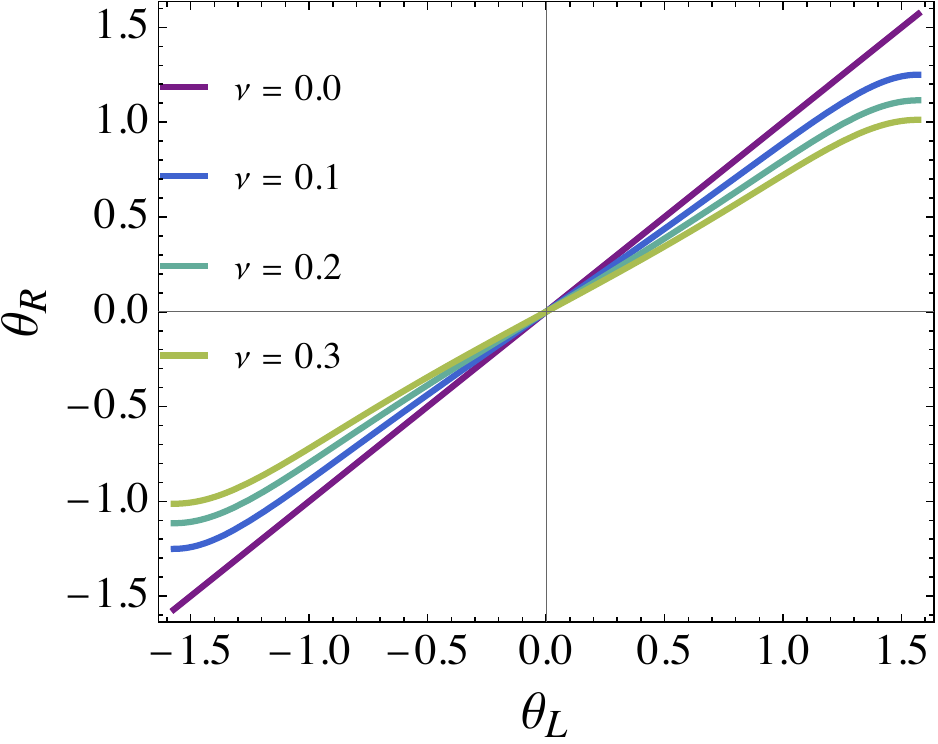}
    \caption{Plot of Eq.~\eqref{eq:tanthetaR} for a geodesic across a spiral / spiral domain wall. }
    \label{fig:tanthetaR}
\end{figure}

Consider a spin texture which interpolates between two distinct spiral phases with wavevectors $\vec q_L = q\hat x$ and $\vec q_R = q \hat y$ as $x$ goes from $-\infty$ to $+\infty$. This could capture a domain wall in a chiral magnet which has multiple degenerate ground state spiral solutions. Concretely, consider 
\begin{equation}\label{eq:spiral_spiral}
\begin{aligned}
    \vec S(\vec x) &= (0,\sin \vec q(x)\cdot \vec x, \cos \vec q(x)\cdot \vec x)\\
    \vec q(x) &= \begin{cases}
        \frac{2\pi}{\lspin}\hat x & x\to -\infty \\ 
        \frac{2\pi}{\lspin}\hat y & x \to + \infty 
    \end{cases}
\end{aligned}
\end{equation}
as shown, for instance, in Fig.~\ref{fig:ferro_spiral}. The metric far to the left and far to the right takes the form
\begin{equation}
\begin{aligned}
    g_L^{ij} &= \delta^{ij} -\frac{\pi^2l_C^2}{\lspin^2}\delta^{ix}\delta^{jx}\\
    g_R^{ij} &= \delta^{ij} -\frac{\pi^2l_C^2}{\lspin^2}\delta^{iy}\delta^{jy}
\end{aligned}
\end{equation}
while $V(\vec x) = h^2/2m\lspin^2$ in both cases. Therefore we can ignore the effect of $V(\vec x)$ for the asymptotic behavior of geodesics and proceed by studying the conserved energy $E = \frac12mg_{ij} \dot x^i\dot x^j.$
In the asymptotic regions, we have 
\begin{equation}
   \begin{cases}  (1+(\pi l_C/\lspin)^2)\dot x^2 + \dot y^2 = C &  x\to -\infty \\
   (1+(\pi l_C/\lspin)^2)\dot y^2 + \dot x^2 = C & x \to +\infty 
   \end{cases}
\end{equation}
with $C$ a constant. 
Assuming the metric globally only depends on $x$, the $y$ momentum $p_y = mg_{yi} \dot x^i$ is conserved. 
In the asymptotic regions, we have 
\begin{equation}
    \begin{cases}
        p_y = m \dot y & x \to -\infty \\
        p_y = m(1+(\pi l_C/\lspin)^2) \dot y & x \to +\infty. 
    \end{cases}
\end{equation}
Let us parameterize
\begin{equation}
    (\dot x, \dot y) = \begin{cases}
        v_L (\cos\theta_L,\sin\theta_L) & x\to -\infty\\
        v_R (\cos\theta_R,\sin\theta_R) & x\to +\infty 
    \end{cases}
\end{equation}
in the asymptotic regions, and for convenience let
\begin{equation}
    \nu =(\pi l_C/\lspin)^2.
\end{equation}
Solving the system of equations yields 
\begin{equation}\label{eq:tanthetaR}
 \tan\theta_R= \frac{\tan\theta_L}{\sqrt{(1+\nu)^3 + \nu(1+\nu) \tan^2\theta_L}}.
\end{equation}
This relation is plotted in Fig.~\ref{fig:tanthetaR}.
To leading order and at small angles, we have
\begin{equation}
    \theta_R - \theta_L \approx -\frac32\theta_L \frac{\pi^2l_C^2}{\lspin^2}. 
\end{equation}

\section{Summary of Riemannian geometry}\label{app:riemann}
In this section, we summarize the results from Riemannian geometry used in the main text. For a thorough description of the Riemannian geometry, see, for example, Ref.~\cite{nakahara_geometry_2018, kobayashi_differential_2019}.
\subsection{Basics of Riemannian geometry}
Consider $d$-dimensional Riemannian manifold $M$. The line element on $M$ defines the metric $g_{ij}$ as 
\begin{align}
    \dd{l}^2 &= g_{ij}\dd{x}^i \dd{x}^j,
\end{align}
where $x^i (i=1, \dots, d)$ is the (local) coordinate. The metric defines the inner product between vectors $A$ and $B$, which we denote as $g(A, B)\equiv g_{ij}A^i B^j$.
We can define the Levi-Civita connection associated with the metric $g_{ij}$. The corresponding Christoffel symbol is given by 
\begin{align}
    \Gamma^i_{jk} &= \frac{g^{im}}{2}\qty(\partial_k g_{mj} + \partial_j g_{mk} - \partial_m g_{jk}),
\end{align}
where $\partial_i = \pdv*{x^i}$. 
With the Christoffel symbol, we can define the covariant derivative of vectors. For a vector $A$, its covariant derivative in $i$-direction, $\nabla_i A$, is given by 
\begin{align}
    (\nabla_i A)^j = \partial_i A^j + \Gamma^j_{ik}A^k.
\end{align}

The Riemann curvature tensor $R^{i}_{~jkl}$ is defined as follows. We denote the basis vector in $i$-direction as $e_i$. $e_i$ is what is identified with the partial derivative $\partial_i$ in mathematics literature. $R^i_{~jkl}$ is then defined through the following relation:
\begin{align}
    \comm{\nabla_k}{\nabla_l} e_j &= R^i_{~jkl}e_i.  
\end{align}
We can show $R^i_{~jkl}$ is written with the Christoffel symbol as 
\begin{align}
    R^i_{~jkl} &= \partial_k \Gamma^i_{lj} - \partial_l \Gamma^i_{kj} + \Gamma^i_{km}\Gamma^m_{lj} - \Gamma^i_{lm}\Gamma^m_{kj}.
\end{align}

\subsection{Gauss-Bonnet theorem}
Here we assume $d=2$. Gauss-Bonnet theorem relates a closed loop on a two-dimensional curved surface to the Gaussian curvature enclosed by the loop. Consider a closed loop $C$, parameterized by its arclength $s$ as $x^i(s)$. $C$ is parametrized so that the vector $\dot{x}\equiv \dv*{x}{s}$ rotates in counterclockwise. $C$ can have vertices with an exterior angle $\phi_i$. Then the Gauss-Bonnet theorem states that the following relation holds for a closed loop $C$:
\begin{align}
    \int_C k_g \dd{s} + \int_S \kappa \sqrt{\det g_{ij}} \dd^2x + \sum_i \phi_i = 2\pi\chi(S).
\end{align}
Here, $S$ is the region enclosed by $C$ and $\chi(S)$ is the Euler characteristic. $\kappa$ is the Gaussian curvature, given by
\begin{align}
    \kappa &= \frac{R_{xyxy}}{\det g_{ij}},
\end{align}
where $R_{ijkl}=g_{im}R^m_{~jkl}$. $k_g$ is the geodesic curvature of the curve $C$, defined as 
\begin{align}
    k_g = g(\nabla_{s}\dot{x}, a).
\end{align}
Here, $\dot{x}=\dv*{x}{s}$ is the tangent vector along the curve $C$. Since $s$ is the arclength, $\dot{x}$ is a unit vector: $g(\dot{x}, \dot{x})=1$. $a$ is a unit vector orthogonal to $\dot{x}$, satisyfing $g(\dot{x}, a) = 0$, $g(a, a)=1$. As we consider a two-dimensional surface, $a$ is unique up to sign, and we fix the sign so that $(\dot{x}, a)$ forms a local right-handed basis. $(\nabla_s \dot{x})^i = \ddot{x}^i + \Gamma^i_{jk}\dot{x}^j\dot{x}^k$ is the covariant derivative of $\dot{x}$ along the curve, with $\ddot{x}^i = \dv*[2]{x^i}{s}$. Since $\nabla_s \dot{x}$ vanishes when $C$ is a geodesic, $k_g$ also vanishes for a geodesic. Therefore, $k_g$ measures the deviation of the curve from geodesics.

When $C$ is a polygon made of geodesics, $\chi(S)=1$ and $k_g=0$. Therefore, it follows
\begin{align}
    \int_S \kappa \sqrt{\det g_{ij}} \dd^2x + \sum_i \phi_i = 2\pi.
\end{align}
This is what we used in the main text.

\section{Relation to the quantum geometry} \label{app:quantum_Reimannian_geometry}
In this section, we briefly comment on the relation between the emergent curved space and the quantum geometry. 

As discussed in the main text, the emergent curved space is characterized by the effective metric given by $g_{ij} = \delta_{ij} + \lp^2 G_{ij}$, with $G_{ij}$ the quantum metric. 
On the other hand, the quantum metric $G_{ij}$ defines the Riemannian geometry of the Hilbert space, which we here call \textit{quantum Riemannian geometry}. In the same way as the standard Riemannian geometry, we can define the Christoffel symbol and the Riemann curvature as:
\begin{align}
    \gamma^i_{jk} &= \frac{(G^{-1})^{im}}{2}\qty(\partial_k G_{mj} + \partial_j G_{mk} - \partial_m G_{jk}), \\
    r^i_{~jkl} &= \partial_k \gamma^i_{jl} - \partial_l \gamma^i_{kj} + \gamma^i_{km}\gamma^m_{lj} - \gamma^i_{lm}\gamma^m_{kj}.
\end{align}
$r^i_{~jkl}$ represents the curvature of the Hilbert space, and we may call the quantum Riemann curvature tensor. We assume here that $G_{ij}$ is positive definite so that we can define its inverse $G^{-1}$.

In this case, the Riemann curvature of the emergent space, $R_{ijkl}$, is directly related to the quantum Riemann curvature, $r_{ijkl}$: 
\begin{align}
    R^i_{~jkl} &= \lp^2 r^i_{~jkl}.
\end{align}

\bibliography{references_zotero,references}

\begin{thebibliography}{52}%
\makeatletter
\providecommand \@ifxundefined [1]{%
 \@ifx{#1\undefined}
}%
\providecommand \@ifnum [1]{%
 \ifnum #1\expandafter \@firstoftwo
 \else \expandafter \@secondoftwo
 \fi
}%
\providecommand \@ifx [1]{%
 \ifx #1\expandafter \@firstoftwo
 \else \expandafter \@secondoftwo
 \fi
}%
\providecommand \natexlab [1]{#1}%
\providecommand \enquote  [1]{``#1''}%
\providecommand \bibnamefont  [1]{#1}%
\providecommand \bibfnamefont [1]{#1}%
\providecommand \citenamefont [1]{#1}%
\providecommand \href@noop [0]{\@secondoftwo}%
\providecommand \href [0]{\begingroup \@sanitize@url \@href}%
\providecommand \@href[1]{\@@startlink{#1}\@@href}%
\providecommand \@@href[1]{\endgroup#1\@@endlink}%
\providecommand \@sanitize@url [0]{\catcode `\\12\catcode `\$12\catcode
  `\&12\catcode `\#12\catcode `\^12\catcode `\_12\catcode `\%12\relax}%
\providecommand \@@startlink[1]{}%
\providecommand \@@endlink[0]{}%
\providecommand \url  [0]{\begingroup\@sanitize@url \@url }%
\providecommand \@url [1]{\endgroup\@href {#1}{\urlprefix }}%
\providecommand \urlprefix  [0]{URL }%
\providecommand \Eprint [0]{\href }%
\providecommand \doibase [0]{https://doi.org/}%
\providecommand \selectlanguage [0]{\@gobble}%
\providecommand \bibinfo  [0]{\@secondoftwo}%
\providecommand \bibfield  [0]{\@secondoftwo}%
\providecommand \translation [1]{[#1]}%
\providecommand \BibitemOpen [0]{}%
\providecommand \bibitemStop [0]{}%
\providecommand \bibitemNoStop [0]{.\EOS\space}%
\providecommand \EOS [0]{\spacefactor3000\relax}%
\providecommand \BibitemShut  [1]{\csname bibitem#1\endcsname}%
\let\auto@bib@innerbib\@empty
\bibitem [{\citenamefont {Sundaram}\ and\ \citenamefont
  {Niu}(1999)}]{sundaram_wave-packet_1999}%
  \BibitemOpen
  \bibfield  {author} {\bibinfo {author} {\bibfnamefont {G.}~\bibnamefont
  {Sundaram}}\ and\ \bibinfo {author} {\bibfnamefont {Q.}~\bibnamefont {Niu}},\
  }\bibfield  {title} {\bibinfo {title} {Wave-packet dynamics in slowly
  perturbed crystals: {Gradient} corrections and {Berry}-phase effects},\
  }\href {https://doi.org/10.1103/PhysRevB.59.14915} {\bibfield  {journal}
  {\bibinfo  {journal} {Physical Review B}\ }\textbf {\bibinfo {volume} {59}},\
  \bibinfo {pages} {14915} (\bibinfo {year} {1999})},\ \bibinfo {note}
  {publisher: American Physical Society}\BibitemShut {NoStop}%
\bibitem [{\citenamefont {Volovik}(1987)}]{volovik_linear_1987}%
  \BibitemOpen
  \bibfield  {author} {\bibinfo {author} {\bibfnamefont {G.~E.}\ \bibnamefont
  {Volovik}},\ }\bibfield  {title} {\bibinfo {title} {Linear momentum in
  ferromagnets},\ }\href {https://doi.org/10.1088/0022-3719/20/7/003}
  {\bibfield  {journal} {\bibinfo  {journal} {Journal of Physics C: Solid State
  Physics}\ }\textbf {\bibinfo {volume} {20}},\ \bibinfo {pages} {L83}
  (\bibinfo {year} {1987})}\BibitemShut {NoStop}%
\bibitem [{\citenamefont {Loss}\ \emph {et~al.}(1990)\citenamefont {Loss},
  \citenamefont {Goldbart},\ and\ \citenamefont {Balatsky}}]{loss_berrys_1990}%
  \BibitemOpen
  \bibfield  {author} {\bibinfo {author} {\bibfnamefont {D.}~\bibnamefont
  {Loss}}, \bibinfo {author} {\bibfnamefont {P.}~\bibnamefont {Goldbart}},\
  and\ \bibinfo {author} {\bibfnamefont {A.~V.}\ \bibnamefont {Balatsky}},\
  }\bibfield  {title} {\bibinfo {title} {Berry’s phase and persistent charge
  and spin currents in textured mesoscopic rings},\ }\href
  {https://doi.org/10.1103/PhysRevLett.65.1655} {\bibfield  {journal} {\bibinfo
   {journal} {Physical Review Letters}\ }\textbf {\bibinfo {volume} {65}},\
  \bibinfo {pages} {1655} (\bibinfo {year} {1990})}\BibitemShut {NoStop}%
\bibitem [{\citenamefont {Ye}\ \emph {et~al.}(1999)\citenamefont {Ye},
  \citenamefont {Kim}, \citenamefont {Millis}, \citenamefont {Shraiman},
  \citenamefont {Majumdar},\ and\ \citenamefont {Tešanović}}]{ye_berry_1999}%
  \BibitemOpen
  \bibfield  {author} {\bibinfo {author} {\bibfnamefont {J.}~\bibnamefont
  {Ye}}, \bibinfo {author} {\bibfnamefont {Y.~B.}\ \bibnamefont {Kim}},
  \bibinfo {author} {\bibfnamefont {A.~J.}\ \bibnamefont {Millis}}, \bibinfo
  {author} {\bibfnamefont {B.~I.}\ \bibnamefont {Shraiman}}, \bibinfo {author}
  {\bibfnamefont {P.}~\bibnamefont {Majumdar}},\ and\ \bibinfo {author}
  {\bibfnamefont {Z.}~\bibnamefont {Tešanović}},\ }\bibfield  {title}
  {\bibinfo {title} {Berry {Phase} {Theory} of the {Anomalous} {Hall} {Effect}:
  {Application} to {Colossal} {Magnetoresistance} {Manganites}},\ }\href
  {https://doi.org/10.1103/PhysRevLett.83.3737} {\bibfield  {journal} {\bibinfo
   {journal} {Physical Review Letters}\ }\textbf {\bibinfo {volume} {83}},\
  \bibinfo {pages} {3737} (\bibinfo {year} {1999})}\BibitemShut {NoStop}%
\bibitem [{\citenamefont {Taguchi}\ \emph {et~al.}(2001)\citenamefont
  {Taguchi}, \citenamefont {Oohara}, \citenamefont {Yoshizawa}, \citenamefont
  {Nagaosa},\ and\ \citenamefont {Tokura}}]{taguchi_spin_2001}%
  \BibitemOpen
  \bibfield  {author} {\bibinfo {author} {\bibfnamefont {Y.}~\bibnamefont
  {Taguchi}}, \bibinfo {author} {\bibfnamefont {Y.}~\bibnamefont {Oohara}},
  \bibinfo {author} {\bibfnamefont {H.}~\bibnamefont {Yoshizawa}}, \bibinfo
  {author} {\bibfnamefont {N.}~\bibnamefont {Nagaosa}},\ and\ \bibinfo {author}
  {\bibfnamefont {Y.}~\bibnamefont {Tokura}},\ }\bibfield  {title} {\bibinfo
  {title} {Spin {Chirality}, {Berry} {Phase}, and {Anomalous} {Hall} {Effect}
  in a {Frustrated} {Ferromagnet}},\ }\href
  {https://doi.org/10.1126/science.1058161} {\bibfield  {journal} {\bibinfo
  {journal} {Science}\ }\textbf {\bibinfo {volume} {291}},\ \bibinfo {pages}
  {2573} (\bibinfo {year} {2001})}\BibitemShut {NoStop}%
\bibitem [{\citenamefont {Onoda}\ \emph {et~al.}(2004)\citenamefont {Onoda},
  \citenamefont {Tatara},\ and\ \citenamefont
  {Nagaosa}}]{onoda_anomalous_2004}%
  \BibitemOpen
  \bibfield  {author} {\bibinfo {author} {\bibfnamefont {M.}~\bibnamefont
  {Onoda}}, \bibinfo {author} {\bibfnamefont {G.}~\bibnamefont {Tatara}},\ and\
  \bibinfo {author} {\bibfnamefont {N.}~\bibnamefont {Nagaosa}},\ }\bibfield
  {title} {\bibinfo {title} {Anomalous {Hall} {Effect} and {Skyrmion} {Number}
  in {Real} and {Momentum} {Spaces}},\ }\href
  {https://doi.org/10.1143/JPSJ.73.2624} {\bibfield  {journal} {\bibinfo
  {journal} {Journal of the Physical Society of Japan}\ }\textbf {\bibinfo
  {volume} {73}},\ \bibinfo {pages} {2624} (\bibinfo {year}
  {2004})}\BibitemShut {NoStop}%
\bibitem [{\citenamefont {Neubauer}\ \emph {et~al.}(2009)\citenamefont
  {Neubauer}, \citenamefont {Pfleiderer}, \citenamefont {Binz}, \citenamefont
  {Rosch}, \citenamefont {Ritz}, \citenamefont {Niklowitz},\ and\ \citenamefont
  {Böni}}]{neubauer_topological_2009}%
  \BibitemOpen
  \bibfield  {author} {\bibinfo {author} {\bibfnamefont {A.}~\bibnamefont
  {Neubauer}}, \bibinfo {author} {\bibfnamefont {C.}~\bibnamefont
  {Pfleiderer}}, \bibinfo {author} {\bibfnamefont {B.}~\bibnamefont {Binz}},
  \bibinfo {author} {\bibfnamefont {A.}~\bibnamefont {Rosch}}, \bibinfo
  {author} {\bibfnamefont {R.}~\bibnamefont {Ritz}}, \bibinfo {author}
  {\bibfnamefont {P.~G.}\ \bibnamefont {Niklowitz}},\ and\ \bibinfo {author}
  {\bibfnamefont {P.}~\bibnamefont {Böni}},\ }\bibfield  {title} {\bibinfo
  {title} {Topological {Hall} {Effect} in the \${A}\$ {Phase} of {MnSi}},\
  }\href {https://doi.org/10.1103/PhysRevLett.102.186602} {\bibfield  {journal}
  {\bibinfo  {journal} {Physical Review Letters}\ }\textbf {\bibinfo {volume}
  {102}},\ \bibinfo {pages} {186602} (\bibinfo {year} {2009})},\ \bibinfo
  {note} {publisher: American Physical Society}\BibitemShut {NoStop}%
\bibitem [{\citenamefont {Ohgushi}\ \emph {et~al.}(2000)\citenamefont
  {Ohgushi}, \citenamefont {Murakami},\ and\ \citenamefont
  {Nagaosa}}]{ohgushi_spin_2000}%
  \BibitemOpen
  \bibfield  {author} {\bibinfo {author} {\bibfnamefont {K.}~\bibnamefont
  {Ohgushi}}, \bibinfo {author} {\bibfnamefont {S.}~\bibnamefont {Murakami}},\
  and\ \bibinfo {author} {\bibfnamefont {N.}~\bibnamefont {Nagaosa}},\
  }\bibfield  {title} {\bibinfo {title} {Spin anisotropy and quantum {Hall}
  effect in the kagom{\textbackslash}'e lattice: {Chiral} spin state based on a
  ferromagnet},\ }\href {https://doi.org/10.1103/PhysRevB.62.R6065} {\bibfield
  {journal} {\bibinfo  {journal} {Physical Review B}\ }\textbf {\bibinfo
  {volume} {62}},\ \bibinfo {pages} {R6065} (\bibinfo {year} {2000})},\
  \bibinfo {note} {publisher: American Physical Society}\BibitemShut {NoStop}%
\bibitem [{\citenamefont {Bruno}\ \emph {et~al.}(2004)\citenamefont {Bruno},
  \citenamefont {Dugaev},\ and\ \citenamefont
  {Taillefumier}}]{bruno_topological_2004}%
  \BibitemOpen
  \bibfield  {author} {\bibinfo {author} {\bibfnamefont {P.}~\bibnamefont
  {Bruno}}, \bibinfo {author} {\bibfnamefont {V.~K.}\ \bibnamefont {Dugaev}},\
  and\ \bibinfo {author} {\bibfnamefont {M.}~\bibnamefont {Taillefumier}},\
  }\bibfield  {title} {\bibinfo {title} {Topological {Hall} {Effect} and
  {Berry} {Phase} in {Magnetic} {Nanostructures}},\ }\href
  {https://doi.org/10.1103/PhysRevLett.93.096806} {\bibfield  {journal}
  {\bibinfo  {journal} {Physical Review Letters}\ }\textbf {\bibinfo {volume}
  {93}},\ \bibinfo {pages} {096806} (\bibinfo {year} {2004})},\ \bibinfo {note}
  {publisher: American Physical Society}\BibitemShut {NoStop}%
\bibitem [{\citenamefont {Hamamoto}\ \emph {et~al.}(2015)\citenamefont
  {Hamamoto}, \citenamefont {Ezawa},\ and\ \citenamefont
  {Nagaosa}}]{hamamoto_quantized_2015}%
  \BibitemOpen
  \bibfield  {author} {\bibinfo {author} {\bibfnamefont {K.}~\bibnamefont
  {Hamamoto}}, \bibinfo {author} {\bibfnamefont {M.}~\bibnamefont {Ezawa}},\
  and\ \bibinfo {author} {\bibfnamefont {N.}~\bibnamefont {Nagaosa}},\
  }\bibfield  {title} {\bibinfo {title} {Quantized topological {Hall} effect in
  skyrmion crystal},\ }\href {https://doi.org/10.1103/PhysRevB.92.115417}
  {\bibfield  {journal} {\bibinfo  {journal} {Physical Review B}\ }\textbf
  {\bibinfo {volume} {92}},\ \bibinfo {pages} {115417} (\bibinfo {year}
  {2015})},\ \bibinfo {note} {publisher: American Physical Society}\BibitemShut
  {NoStop}%
\bibitem [{\citenamefont {Addison}\ \emph {et~al.}(2024)\citenamefont
  {Addison}, \citenamefont {Keyes},\ and\ \citenamefont
  {Randeria}}]{addison_anomalous_2024}%
  \BibitemOpen
  \bibfield  {author} {\bibinfo {author} {\bibfnamefont {Z.}~\bibnamefont
  {Addison}}, \bibinfo {author} {\bibfnamefont {L.}~\bibnamefont {Keyes}},\
  and\ \bibinfo {author} {\bibfnamefont {M.}~\bibnamefont {Randeria}},\ }\href
  {https://doi.org/10.48550/arXiv.2409.04376} {\bibinfo {title} {Anomalous and
  {Topological} {Hall} {Effects} with {Phase}-{Space} {Berry} {Curvatures}:
  {Electric}, {Thermal}, and {Thermoelectric} {Transport} in {Magnets}}}
  (\bibinfo {year} {2024}),\ \bibinfo {note} {arXiv:2409.04376
  [cond-mat]}\BibitemShut {NoStop}%
\bibitem [{\citenamefont {Park}\ \emph {et~al.}(2025)\citenamefont {Park},
  \citenamefont {Huang}, \citenamefont {Savary},\ and\ \citenamefont
  {Balents}}]{park_quantum_2025}%
  \BibitemOpen
  \bibfield  {author} {\bibinfo {author} {\bibfnamefont {T.}~\bibnamefont
  {Park}}, \bibinfo {author} {\bibfnamefont {X.}~\bibnamefont {Huang}},
  \bibinfo {author} {\bibfnamefont {L.}~\bibnamefont {Savary}},\ and\ \bibinfo
  {author} {\bibfnamefont {L.}~\bibnamefont {Balents}},\ }\href
  {https://doi.org/10.48550/arXiv.2504.10447} {\bibinfo {title} {Quantum
  geometry from the {Moyal} product: quantum kinetic equation and non-linear
  response}} (\bibinfo {year} {2025}),\ \bibinfo {note} {arXiv:2504.10447
  [cond-mat]}\BibitemShut {NoStop}%
\bibitem [{\citenamefont {Paul}\ \emph {et~al.}(2023)\citenamefont {Paul},
  \citenamefont {Zhang},\ and\ \citenamefont {Fu}}]{paul_giant_2023}%
  \BibitemOpen
  \bibfield  {author} {\bibinfo {author} {\bibfnamefont {N.}~\bibnamefont
  {Paul}}, \bibinfo {author} {\bibfnamefont {Y.}~\bibnamefont {Zhang}},\ and\
  \bibinfo {author} {\bibfnamefont {L.}~\bibnamefont {Fu}},\ }\bibfield
  {title} {\bibinfo {title} {Giant proximity exchange and flat {Chern} band in
  {2D} magnet-semiconductor heterostructures},\ }\href
  {https://doi.org/10.1126/sciadv.abn1401} {\bibfield  {journal} {\bibinfo
  {journal} {Science Advances}\ }\textbf {\bibinfo {volume} {9}},\ \bibinfo
  {pages} {eabn1401} (\bibinfo {year} {2023})},\ \bibinfo {note} {publisher:
  American Association for the Advancement of Science}\BibitemShut {NoStop}%
\bibitem [{\citenamefont {Schneider}\ \emph {et~al.}(1992)\citenamefont
  {Schneider}, \citenamefont {Ehlers},\ and\ \citenamefont
  {Falco}}]{schneider_gravitational_1992}%
  \BibitemOpen
  \bibfield  {author} {\bibinfo {author} {\bibfnamefont {P.}~\bibnamefont
  {Schneider}}, \bibinfo {author} {\bibfnamefont {J.}~\bibnamefont {Ehlers}},\
  and\ \bibinfo {author} {\bibfnamefont {E.~E.}\ \bibnamefont {Falco}},\ }\href
  {https://doi.org/10.1007/978-3-662-03758-4} {\emph {\bibinfo {title}
  {Gravitational {Lenses}}}},\ edited by\ \bibinfo {editor} {\bibfnamefont
  {I.}~\bibnamefont {Appenzeller}}, \bibinfo {editor} {\bibfnamefont
  {G.}~\bibnamefont {Börner}}, \bibinfo {editor} {\bibfnamefont
  {M.}~\bibnamefont {Harwit}}, \bibinfo {editor} {\bibfnamefont
  {R.}~\bibnamefont {Kippenhahn}}, \bibinfo {editor} {\bibfnamefont
  {J.}~\bibnamefont {Lequeux}}, \bibinfo {editor} {\bibfnamefont {P.~A.}\
  \bibnamefont {Strittmatter}},\ and\ \bibinfo {editor} {\bibfnamefont
  {V.}~\bibnamefont {Trimble}},\ Astronomy and {Astrophysics} {Library}\
  (\bibinfo  {publisher} {Springer Berlin Heidelberg},\ \bibinfo {address}
  {Berlin, Heidelberg},\ \bibinfo {year} {1992})\BibitemShut {NoStop}%
\bibitem [{\citenamefont {Fujita}\ \emph {et~al.}(2011)\citenamefont {Fujita},
  \citenamefont {Jalil}, \citenamefont {Tan},\ and\ \citenamefont
  {Murakami}}]{fujita_gauge_2011}%
  \BibitemOpen
  \bibfield  {author} {\bibinfo {author} {\bibfnamefont {T.}~\bibnamefont
  {Fujita}}, \bibinfo {author} {\bibfnamefont {M.~B.~A.}\ \bibnamefont
  {Jalil}}, \bibinfo {author} {\bibfnamefont {S.~G.}\ \bibnamefont {Tan}},\
  and\ \bibinfo {author} {\bibfnamefont {S.}~\bibnamefont {Murakami}},\
  }\bibfield  {title} {\bibinfo {title} {Gauge fields in spintronics},\ }\href
  {https://doi.org/10.1063/1.3665219} {\bibfield  {journal} {\bibinfo
  {journal} {Journal of Applied Physics}\ }\textbf {\bibinfo {volume} {110}},\
  \bibinfo {pages} {121301} (\bibinfo {year} {2011})}\BibitemShut {NoStop}%
\bibitem [{\citenamefont {Tan}\ \emph {et~al.}(2020)\citenamefont {Tan},
  \citenamefont {Chen}, \citenamefont {Ho}, \citenamefont {Huang},
  \citenamefont {Jalil}, \citenamefont {Chang},\ and\ \citenamefont
  {Murakami}}]{tan_yangmills_2020}%
  \BibitemOpen
  \bibfield  {author} {\bibinfo {author} {\bibfnamefont {S.~G.}\ \bibnamefont
  {Tan}}, \bibinfo {author} {\bibfnamefont {S.-H.}\ \bibnamefont {Chen}},
  \bibinfo {author} {\bibfnamefont {C.~S.}\ \bibnamefont {Ho}}, \bibinfo
  {author} {\bibfnamefont {C.-C.}\ \bibnamefont {Huang}}, \bibinfo {author}
  {\bibfnamefont {M.~B.~A.}\ \bibnamefont {Jalil}}, \bibinfo {author}
  {\bibfnamefont {C.~R.}\ \bibnamefont {Chang}},\ and\ \bibinfo {author}
  {\bibfnamefont {S.}~\bibnamefont {Murakami}},\ }\bibfield  {title} {\bibinfo
  {title} {Yang–{Mills} physics in spintronics},\ }\href
  {https://doi.org/10.1016/j.physrep.2020.08.002} {\bibfield  {journal}
  {\bibinfo  {journal} {Physics Reports}\ }\bibinfo {series} {Yang-{Mills}
  physics in spintronics},\ \textbf {\bibinfo {volume} {882}},\ \bibinfo
  {pages} {1} (\bibinfo {year} {2020})}\BibitemShut {NoStop}%
\bibitem [{\citenamefont {De~Gennes}(1960)}]{de_gennes_effects_1960}%
  \BibitemOpen
  \bibfield  {author} {\bibinfo {author} {\bibfnamefont {P.~G.}\ \bibnamefont
  {De~Gennes}},\ }\bibfield  {title} {\bibinfo {title} {Effects of {Double}
  {Exchange} in {Magnetic} {Crystals}},\ }\href
  {https://doi.org/10.1103/PhysRev.118.141} {\bibfield  {journal} {\bibinfo
  {journal} {Physical Review}\ }\textbf {\bibinfo {volume} {118}},\ \bibinfo
  {pages} {141} (\bibinfo {year} {1960})}\BibitemShut {NoStop}%
\bibitem [{\citenamefont {Ohnishi}\ and\ \citenamefont
  {Nagaosa}(2007)}]{ohnishi_adiabatic_2007}%
  \BibitemOpen
  \bibfield  {author} {\bibinfo {author} {\bibfnamefont {T.}~\bibnamefont
  {Ohnishi}}\ and\ \bibinfo {author} {\bibfnamefont {N.}~\bibnamefont
  {Nagaosa}},\ }\bibfield  {title} {\bibinfo {title} {Adiabatic {Approximation}
  for {Path} {Integrals} and {Geometrical} {Potentials}},\ }\href
  {https://doi.org/10.1143/JPSJ.76.015003} {\bibfield  {journal} {\bibinfo
  {journal} {Journal of the Physical Society of Japan}\ }\textbf {\bibinfo
  {volume} {76}},\ \bibinfo {pages} {015003} (\bibinfo {year} {2007})},\
  \bibinfo {note} {publisher: The Physical Society of Japan}\BibitemShut
  {NoStop}%
\bibitem [{\citenamefont {Aharonov}(1992)}]{aharonov_origin_1992}%
  \BibitemOpen
  \bibfield  {author} {\bibinfo {author} {\bibfnamefont {Y.}~\bibnamefont
  {Aharonov}},\ }\bibfield  {title} {\bibinfo {title} {Origin of the geometric
  forces accompanying {Berry}’s geometric potentials},\ }\href
  {https://doi.org/10.1103/PhysRevLett.69.3593} {\bibfield  {journal} {\bibinfo
   {journal} {Physical Review Letters}\ }\textbf {\bibinfo {volume} {69}},\
  \bibinfo {pages} {3593} (\bibinfo {year} {1992})}\BibitemShut {NoStop}%
\bibitem [{\citenamefont {Berry}\ and\ \citenamefont
  {Robbins}(1997)}]{berry_classical_1997}%
  \BibitemOpen
  \bibfield  {author} {\bibinfo {author} {\bibfnamefont {M.~V.}\ \bibnamefont
  {Berry}}\ and\ \bibinfo {author} {\bibfnamefont {J.~M.}\ \bibnamefont
  {Robbins}},\ }\bibfield  {title} {\bibinfo {title} {Classical geometric
  forces of reaction: an exactly solvable model},\ }\href
  {https://doi.org/10.1098/rspa.1993.0126} {\bibfield  {journal} {\bibinfo
  {journal} {Proceedings of the Royal Society of London. Series A: Mathematical
  and Physical Sciences}\ }\textbf {\bibinfo {volume} {442}},\ \bibinfo {pages}
  {641} (\bibinfo {year} {1997})},\ \bibinfo {note} {publisher: Royal
  Society}\BibitemShut {NoStop}%
\bibitem [{Note1()}]{Note1}%
  \BibitemOpen
  \bibinfo {note} {We note that the Hamiltonian for a particle in curved space
  can commonly take a different form, but we can show that the two formulations
  are equivalent up to $\order {l_C^2\lambda _s^{-4}}$. See SM for more
  details.}\BibitemShut {Stop}%
\bibitem [{\citenamefont {Deser}\ \emph {et~al.}(1984)\citenamefont {Deser},
  \citenamefont {Jackiw},\ and\ \citenamefont {'t~Hooft}}]{Deser1984Jan}%
  \BibitemOpen
  \bibfield  {author} {\bibinfo {author} {\bibfnamefont {S.}~\bibnamefont
  {Deser}}, \bibinfo {author} {\bibfnamefont {R.}~\bibnamefont {Jackiw}},\ and\
  \bibinfo {author} {\bibfnamefont {G.}~\bibnamefont {'t~Hooft}},\ }\bibfield
  {title} {\bibinfo {title} {{Three-dimensional Einstein gravity: Dynamics of
  flat space}},\ }\href {https://doi.org/10.1016/0003-4916(84)90085-X}
  {\bibfield  {journal} {\bibinfo  {journal} {Ann. Phys.}\ }\textbf {\bibinfo
  {volume} {152}},\ \bibinfo {pages} {220} (\bibinfo {year}
  {1984})}\BibitemShut {NoStop}%
\bibitem [{\citenamefont {Nakahara}(2018)}]{nakahara_geometry_2018}%
  \BibitemOpen
  \bibfield  {author} {\bibinfo {author} {\bibfnamefont {M.}~\bibnamefont
  {Nakahara}},\ }\href {https://doi.org/10.1201/9781315275826} {\emph {\bibinfo
  {title} {Geometry, {Topology} and {Physics}}}},\ \bibinfo {edition} {2nd}\
  ed.\ (\bibinfo  {publisher} {CRC Press},\ \bibinfo {address} {Boca Raton},\
  \bibinfo {year} {2018})\BibitemShut {NoStop}%
\bibitem [{\citenamefont {Kobayashi}(2019)}]{kobayashi_differential_2019}%
  \BibitemOpen
  \bibfield  {author} {\bibinfo {author} {\bibfnamefont {S.}~\bibnamefont
  {Kobayashi}},\ }\href {https://doi.org/10.1007/978-981-15-1739-6} {\emph
  {\bibinfo {title} {Differential {Geometry} of {Curves} and {Surfaces}}}},\
  Springer {Undergraduate} {Mathematics} {Series}\ (\bibinfo  {publisher}
  {Springer Singapore},\ \bibinfo {address} {Singapore},\ \bibinfo {year}
  {2019})\BibitemShut {NoStop}%
\bibitem [{Note2()}]{Note2}%
  \BibitemOpen
  \bibinfo {note} {After completing this work, we found similar relations for
  the true gravitational lensing of light in Ref.~\cite
  {gibbons_applications_2008}.}\BibitemShut {Stop}%
\bibitem [{\citenamefont {Egerton}(2016)}]{egerton_physical_2016}%
  \BibitemOpen
  \bibfield  {author} {\bibinfo {author} {\bibfnamefont {R.}~\bibnamefont
  {Egerton}},\ }\href {https://doi.org/10.1007/978-3-319-39877-8} {\emph
  {\bibinfo {title} {Physical {Principles} of {Electron} {Microscopy}}}}\
  (\bibinfo  {publisher} {Springer International Publishing},\ \bibinfo
  {address} {Cham},\ \bibinfo {year} {2016})\BibitemShut {NoStop}%
\bibitem [{\citenamefont {Avron}\ \emph {et~al.}(1995)\citenamefont {Avron},
  \citenamefont {Seiler},\ and\ \citenamefont {Zograf}}]{avron_viscosity_1995}%
  \BibitemOpen
  \bibfield  {author} {\bibinfo {author} {\bibfnamefont {J.~E.}\ \bibnamefont
  {Avron}}, \bibinfo {author} {\bibfnamefont {R.}~\bibnamefont {Seiler}},\ and\
  \bibinfo {author} {\bibfnamefont {P.~G.}\ \bibnamefont {Zograf}},\ }\bibfield
   {title} {\bibinfo {title} {Viscosity of {Quantum} {Hall} {Fluids}},\ }\href
  {https://doi.org/10.1103/PhysRevLett.75.697} {\bibfield  {journal} {\bibinfo
  {journal} {Physical Review Letters}\ }\textbf {\bibinfo {volume} {75}},\
  \bibinfo {pages} {697} (\bibinfo {year} {1995})}\BibitemShut {NoStop}%
\bibitem [{\citenamefont {Hughes}\ \emph {et~al.}(2011)\citenamefont {Hughes},
  \citenamefont {Leigh},\ and\ \citenamefont
  {Fradkin}}]{hughes_torsional_2011}%
  \BibitemOpen
  \bibfield  {author} {\bibinfo {author} {\bibfnamefont {T.~L.}\ \bibnamefont
  {Hughes}}, \bibinfo {author} {\bibfnamefont {R.~G.}\ \bibnamefont {Leigh}},\
  and\ \bibinfo {author} {\bibfnamefont {E.}~\bibnamefont {Fradkin}},\
  }\bibfield  {title} {\bibinfo {title} {Torsional {Response} and
  {Dissipationless} {Viscosity} in {Topological} {Insulators}},\ }\href
  {https://doi.org/10.1103/PhysRevLett.107.075502} {\bibfield  {journal}
  {\bibinfo  {journal} {Physical Review Letters}\ }\textbf {\bibinfo {volume}
  {107}},\ \bibinfo {pages} {075502} (\bibinfo {year} {2011})},\ \bibinfo
  {note} {publisher: American Physical Society}\BibitemShut {NoStop}%
\bibitem [{\citenamefont {Bradlyn}\ \emph {et~al.}(2012)\citenamefont
  {Bradlyn}, \citenamefont {Goldstein},\ and\ \citenamefont
  {Read}}]{bradlyn_kubo_2012}%
  \BibitemOpen
  \bibfield  {author} {\bibinfo {author} {\bibfnamefont {B.}~\bibnamefont
  {Bradlyn}}, \bibinfo {author} {\bibfnamefont {M.}~\bibnamefont {Goldstein}},\
  and\ \bibinfo {author} {\bibfnamefont {N.}~\bibnamefont {Read}},\ }\bibfield
  {title} {\bibinfo {title} {Kubo formulas for viscosity: {Hall} viscosity,
  {Ward} identities, and the relation with conductivity},\ }\href
  {https://doi.org/10.1103/PhysRevB.86.245309} {\bibfield  {journal} {\bibinfo
  {journal} {Physical Review B}\ }\textbf {\bibinfo {volume} {86}},\ \bibinfo
  {pages} {245309} (\bibinfo {year} {2012})}\BibitemShut {NoStop}%
\bibitem [{\citenamefont {Dong}\ and\ \citenamefont
  {Niu}(2018)}]{dong_geometrodynamics_2018}%
  \BibitemOpen
  \bibfield  {author} {\bibinfo {author} {\bibfnamefont {L.}~\bibnamefont
  {Dong}}\ and\ \bibinfo {author} {\bibfnamefont {Q.}~\bibnamefont {Niu}},\
  }\bibfield  {title} {\bibinfo {title} {Geometrodynamics of electrons in a
  crystal under position and time-dependent deformation},\ }\href
  {https://doi.org/10.1103/PhysRevB.98.115162} {\bibfield  {journal} {\bibinfo
  {journal} {Physical Review B}\ }\textbf {\bibinfo {volume} {98}},\ \bibinfo
  {pages} {115162} (\bibinfo {year} {2018})},\ \bibinfo {note} {publisher:
  American Physical Society}\BibitemShut {NoStop}%
\bibitem [{\citenamefont {Rao}\ and\ \citenamefont
  {Bradlyn}(2020)}]{rao_hall_2020}%
  \BibitemOpen
  \bibfield  {author} {\bibinfo {author} {\bibfnamefont {P.}~\bibnamefont
  {Rao}}\ and\ \bibinfo {author} {\bibfnamefont {B.}~\bibnamefont {Bradlyn}},\
  }\bibfield  {title} {\bibinfo {title} {Hall {Viscosity} in {Quantum}
  {Systems} with {Discrete} {Symmetry}: {Point} {Group} and {Lattice}
  {Anisotropy}},\ }\href {https://doi.org/10.1103/PhysRevX.10.021005}
  {\bibfield  {journal} {\bibinfo  {journal} {Physical Review X}\ }\textbf
  {\bibinfo {volume} {10}},\ \bibinfo {pages} {021005} (\bibinfo {year}
  {2020})}\BibitemShut {NoStop}%
\bibitem [{\citenamefont {Dahlhaus}\ \emph {et~al.}(2010)\citenamefont
  {Dahlhaus}, \citenamefont {Hou}, \citenamefont {Akhmerov},\ and\
  \citenamefont {Beenakker}}]{dahlhaus_geodesic_2010}%
  \BibitemOpen
  \bibfield  {author} {\bibinfo {author} {\bibfnamefont {J.~P.}\ \bibnamefont
  {Dahlhaus}}, \bibinfo {author} {\bibfnamefont {C.-Y.}\ \bibnamefont {Hou}},
  \bibinfo {author} {\bibfnamefont {A.~R.}\ \bibnamefont {Akhmerov}},\ and\
  \bibinfo {author} {\bibfnamefont {C.~W.~J.}\ \bibnamefont {Beenakker}},\
  }\bibfield  {title} {\bibinfo {title} {Geodesic scattering by surface
  deformations of a topological insulator},\ }\href
  {https://doi.org/10.1103/PhysRevB.82.085312} {\bibfield  {journal} {\bibinfo
  {journal} {Physical Review B}\ }\textbf {\bibinfo {volume} {82}},\ \bibinfo
  {pages} {085312} (\bibinfo {year} {2010})}\BibitemShut {NoStop}%
\bibitem [{\citenamefont {Fischer}\ and\ \citenamefont
  {Visser}(2002)}]{fischer_riemannian_2002}%
  \BibitemOpen
  \bibfield  {author} {\bibinfo {author} {\bibfnamefont {U.~R.}\ \bibnamefont
  {Fischer}}\ and\ \bibinfo {author} {\bibfnamefont {M.}~\bibnamefont
  {Visser}},\ }\bibfield  {title} {\bibinfo {title} {Riemannian {Geometry} of
  {Irrotational} {Vortex} {Acoustics}},\ }\href
  {https://doi.org/10.1103/PhysRevLett.88.110201} {\bibfield  {journal}
  {\bibinfo  {journal} {Physical Review Letters}\ }\textbf {\bibinfo {volume}
  {88}},\ \bibinfo {pages} {110201} (\bibinfo {year} {2002})},\ \bibinfo {note}
  {publisher: American Physical Society}\BibitemShut {NoStop}%
\bibitem [{\citenamefont {Volovik}(2016)}]{volovik_black_2016}%
  \BibitemOpen
  \bibfield  {author} {\bibinfo {author} {\bibfnamefont {G.~E.}\ \bibnamefont
  {Volovik}},\ }\bibfield  {title} {\bibinfo {title} {Black hole and hawking
  radiation by type-{II} {Weyl} fermions},\ }\href
  {https://doi.org/10.1134/S0021364016210050} {\bibfield  {journal} {\bibinfo
  {journal} {JETP Letters}\ }\textbf {\bibinfo {volume} {104}},\ \bibinfo
  {pages} {645} (\bibinfo {year} {2016})},\ \bibinfo {note} {company: Springer
  Distributor: Springer Institution: Springer Label: Springer Number: 9
  Publisher: Pleiades Publishing}\BibitemShut {NoStop}%
\bibitem [{\citenamefont {Westström}\ and\ \citenamefont
  {Ojanen}(2017)}]{weststrom_designer_2017}%
  \BibitemOpen
  \bibfield  {author} {\bibinfo {author} {\bibfnamefont {A.}~\bibnamefont
  {Westström}}\ and\ \bibinfo {author} {\bibfnamefont {T.}~\bibnamefont
  {Ojanen}},\ }\bibfield  {title} {\bibinfo {title} {Designer {Curved}-{Space}
  {Geometry} for {Relativistic} {Fermions} in {Weyl} {Metamaterials}},\ }\href
  {https://doi.org/10.1103/PhysRevX.7.041026} {\bibfield  {journal} {\bibinfo
  {journal} {Physical Review X}\ }\textbf {\bibinfo {volume} {7}},\ \bibinfo
  {pages} {041026} (\bibinfo {year} {2017})}\BibitemShut {NoStop}%
\bibitem [{\citenamefont {Liang}\ and\ \citenamefont
  {Ojanen}(2019)}]{liang_curved_2019}%
  \BibitemOpen
  \bibfield  {author} {\bibinfo {author} {\bibfnamefont {L.}~\bibnamefont
  {Liang}}\ and\ \bibinfo {author} {\bibfnamefont {T.}~\bibnamefont {Ojanen}},\
  }\bibfield  {title} {\bibinfo {title} {Curved spacetime theory of
  inhomogeneous {Weyl} materials},\ }\href
  {https://doi.org/10.1103/PhysRevResearch.1.032006} {\bibfield  {journal}
  {\bibinfo  {journal} {Physical Review Research}\ }\textbf {\bibinfo {volume}
  {1}},\ \bibinfo {pages} {032006} (\bibinfo {year} {2019})},\ \bibinfo {note}
  {publisher: American Physical Society}\BibitemShut {NoStop}%
\bibitem [{\citenamefont {Huang}\ \emph {et~al.}(2018)\citenamefont {Huang},
  \citenamefont {Jin},\ and\ \citenamefont {Liu}}]{huang_black-hole_2018}%
  \BibitemOpen
  \bibfield  {author} {\bibinfo {author} {\bibfnamefont {H.}~\bibnamefont
  {Huang}}, \bibinfo {author} {\bibfnamefont {K.-H.}\ \bibnamefont {Jin}},\
  and\ \bibinfo {author} {\bibfnamefont {F.}~\bibnamefont {Liu}},\ }\bibfield
  {title} {\bibinfo {title} {Black-hole horizon in the {Dirac} semimetal {Zn} 2
  {In} 2 {S} 5},\ }\href {https://doi.org/10.1103/PhysRevB.98.121110}
  {\bibfield  {journal} {\bibinfo  {journal} {Physical Review B}\ }\textbf
  {\bibinfo {volume} {98}},\ \bibinfo {pages} {121110} (\bibinfo {year}
  {2018})}\BibitemShut {NoStop}%
\bibitem [{\citenamefont {Zubkov}(2018)}]{zubkov_black_2018}%
  \BibitemOpen
  \bibfield  {author} {\bibinfo {author} {\bibfnamefont {M.~A.}\ \bibnamefont
  {Zubkov}},\ }\bibfield  {title} {\bibinfo {title} {The black hole interior
  and the type {II} {Weyl} fermions},\ }\href
  {https://doi.org/10.1142/S0217732318500475} {\bibfield  {journal} {\bibinfo
  {journal} {Modern Physics Letters A}\ }\textbf {\bibinfo {volume} {33}},\
  \bibinfo {pages} {1850047} (\bibinfo {year} {2018})},\ \bibinfo {note}
  {publisher: World Scientific Publishing Co.}\BibitemShut {Stop}%
\bibitem [{\citenamefont {Zubkov}(2015)}]{zubkov_emergent_2015}%
  \BibitemOpen
  \bibfield  {author} {\bibinfo {author} {\bibfnamefont {M.~A.}\ \bibnamefont
  {Zubkov}},\ }\bibfield  {title} {\bibinfo {title} {Emergent gravity and
  chiral anomaly in {Dirac} semimetals in the presence of dislocations},\
  }\href {https://doi.org/10.1016/j.aop.2015.05.032} {\bibfield  {journal}
  {\bibinfo  {journal} {Annals of Physics}\ }\textbf {\bibinfo {volume}
  {360}},\ \bibinfo {pages} {655} (\bibinfo {year} {2015})}\BibitemShut
  {NoStop}%
\bibitem [{\citenamefont {Guan}\ \emph {et~al.}(2017)\citenamefont {Guan},
  \citenamefont {Yu}, \citenamefont {Liu}, \citenamefont {Liu}, \citenamefont
  {Dong}, \citenamefont {Lu}, \citenamefont {Yao},\ and\ \citenamefont
  {Yang}}]{guan_artificial_2017}%
  \BibitemOpen
  \bibfield  {author} {\bibinfo {author} {\bibfnamefont {S.}~\bibnamefont
  {Guan}}, \bibinfo {author} {\bibfnamefont {Z.-M.}\ \bibnamefont {Yu}},
  \bibinfo {author} {\bibfnamefont {Y.}~\bibnamefont {Liu}}, \bibinfo {author}
  {\bibfnamefont {G.-B.}\ \bibnamefont {Liu}}, \bibinfo {author} {\bibfnamefont
  {L.}~\bibnamefont {Dong}}, \bibinfo {author} {\bibfnamefont {Y.}~\bibnamefont
  {Lu}}, \bibinfo {author} {\bibfnamefont {Y.}~\bibnamefont {Yao}},\ and\
  \bibinfo {author} {\bibfnamefont {S.~A.}\ \bibnamefont {Yang}},\ }\bibfield
  {title} {\bibinfo {title} {Artificial gravity field, astrophysical analogues,
  and topological phase transitions in strained topological semimetals},\
  }\href {https://doi.org/10.1038/s41535-017-0026-7} {\bibfield  {journal}
  {\bibinfo  {journal} {npj Quantum Materials}\ }\textbf {\bibinfo {volume}
  {2}},\ \bibinfo {pages} {23} (\bibinfo {year} {2017})},\ \bibinfo {note}
  {publisher: Nature Publishing Group}\BibitemShut {NoStop}%
\bibitem [{\citenamefont {Nissinen}\ and\ \citenamefont
  {Volovik}(2017)}]{nissinen_type-iii_2017}%
  \BibitemOpen
  \bibfield  {author} {\bibinfo {author} {\bibfnamefont {J.}~\bibnamefont
  {Nissinen}}\ and\ \bibinfo {author} {\bibfnamefont {G.~E.}\ \bibnamefont
  {Volovik}},\ }\bibfield  {title} {\bibinfo {title} {Type-{III} and {IV}
  interacting {Weyl} points},\ }\href
  {https://doi.org/10.1134/S0021364017070013} {\bibfield  {journal} {\bibinfo
  {journal} {JETP Letters}\ }\textbf {\bibinfo {volume} {105}},\ \bibinfo
  {pages} {447} (\bibinfo {year} {2017})}\BibitemShut {NoStop}%
\bibitem [{\citenamefont {Könye}\ \emph {et~al.}(2023)\citenamefont {Könye},
  \citenamefont {Mertens}, \citenamefont {Morice}, \citenamefont {Chernyavsky},
  \citenamefont {Moghaddam}, \citenamefont {van Wezel},\ and\ \citenamefont
  {van~den Brink}}]{konye_anisotropic_2023}%
  \BibitemOpen
  \bibfield  {author} {\bibinfo {author} {\bibfnamefont {V.}~\bibnamefont
  {Könye}}, \bibinfo {author} {\bibfnamefont {L.}~\bibnamefont {Mertens}},
  \bibinfo {author} {\bibfnamefont {C.}~\bibnamefont {Morice}}, \bibinfo
  {author} {\bibfnamefont {D.}~\bibnamefont {Chernyavsky}}, \bibinfo {author}
  {\bibfnamefont {A.~G.}\ \bibnamefont {Moghaddam}}, \bibinfo {author}
  {\bibfnamefont {J.}~\bibnamefont {van Wezel}},\ and\ \bibinfo {author}
  {\bibfnamefont {J.}~\bibnamefont {van~den Brink}},\ }\bibfield  {title}
  {\bibinfo {title} {Anisotropic optics and gravitational lensing of tilted
  {Weyl} fermions},\ }\href {https://doi.org/10.1103/PhysRevB.107.L201406}
  {\bibfield  {journal} {\bibinfo  {journal} {Physical Review B}\ }\textbf
  {\bibinfo {volume} {107}},\ \bibinfo {pages} {L201406} (\bibinfo {year}
  {2023})},\ \bibinfo {note} {publisher: American Physical Society}\BibitemShut
  {NoStop}%
\bibitem [{\citenamefont {Haller}\ \emph {et~al.}(2023)\citenamefont {Haller},
  \citenamefont {Hegde}, \citenamefont {Xu}, \citenamefont {De~Beule},
  \citenamefont {Schmidt},\ and\ \citenamefont {Meng}}]{haller_black_2023}%
  \BibitemOpen
  \bibfield  {author} {\bibinfo {author} {\bibfnamefont {A.}~\bibnamefont
  {Haller}}, \bibinfo {author} {\bibfnamefont {S.}~\bibnamefont {Hegde}},
  \bibinfo {author} {\bibfnamefont {C.}~\bibnamefont {Xu}}, \bibinfo {author}
  {\bibfnamefont {C.}~\bibnamefont {De~Beule}}, \bibinfo {author}
  {\bibfnamefont {T.~L.}\ \bibnamefont {Schmidt}},\ and\ \bibinfo {author}
  {\bibfnamefont {T.}~\bibnamefont {Meng}},\ }\bibfield  {title} {\bibinfo
  {title} {Black hole mirages: {Electron} lensing and {Berry} curvature effects
  in inhomogeneously tilted {Weyl} semimetals},\ }\href
  {https://doi.org/10.21468/SciPostPhys.14.5.119} {\bibfield  {journal}
  {\bibinfo  {journal} {SciPost Physics}\ }\textbf {\bibinfo {volume} {14}},\
  \bibinfo {pages} {119} (\bibinfo {year} {2023})}\BibitemShut {NoStop}%
\bibitem [{\citenamefont {Barnes}\ and\ \citenamefont
  {Maekawa}(2007)}]{barnes_generalization_2007}%
  \BibitemOpen
  \bibfield  {author} {\bibinfo {author} {\bibfnamefont {S.~E.}\ \bibnamefont
  {Barnes}}\ and\ \bibinfo {author} {\bibfnamefont {S.}~\bibnamefont
  {Maekawa}},\ }\bibfield  {title} {\bibinfo {title} {Generalization of
  {Faraday}’s {Law} to {Include} {Nonconservative} {Spin} {Forces}},\ }\href
  {https://doi.org/10.1103/PhysRevLett.98.246601} {\bibfield  {journal}
  {\bibinfo  {journal} {Physical Review Letters}\ }\textbf {\bibinfo {volume}
  {98}},\ \bibinfo {pages} {246601} (\bibinfo {year} {2007})}\BibitemShut
  {NoStop}%
\bibitem [{\citenamefont {Nagaosa}\ and\ \citenamefont
  {Tokura}(2013)}]{nagaosa_topological_2013}%
  \BibitemOpen
  \bibfield  {author} {\bibinfo {author} {\bibfnamefont {N.}~\bibnamefont
  {Nagaosa}}\ and\ \bibinfo {author} {\bibfnamefont {Y.}~\bibnamefont
  {Tokura}},\ }\bibfield  {title} {\bibinfo {title} {Topological properties and
  dynamics of magnetic skyrmions},\ }\href
  {https://doi.org/10.1038/nnano.2013.243} {\bibfield  {journal} {\bibinfo
  {journal} {Nature Nanotechnology}\ }\textbf {\bibinfo {volume} {8}},\
  \bibinfo {pages} {899} (\bibinfo {year} {2013})},\ \bibinfo {note}
  {publisher: Nature Publishing Group}\BibitemShut {NoStop}%
\bibitem [{\citenamefont {Yang}\ \emph {et~al.}(2009)\citenamefont {Yang},
  \citenamefont {Beach}, \citenamefont {Knutson}, \citenamefont {Xiao},
  \citenamefont {Niu}, \citenamefont {Tsoi},\ and\ \citenamefont
  {Erskine}}]{yang_universal_2009}%
  \BibitemOpen
  \bibfield  {author} {\bibinfo {author} {\bibfnamefont {S.~A.}\ \bibnamefont
  {Yang}}, \bibinfo {author} {\bibfnamefont {G.~S.~D.}\ \bibnamefont {Beach}},
  \bibinfo {author} {\bibfnamefont {C.}~\bibnamefont {Knutson}}, \bibinfo
  {author} {\bibfnamefont {D.}~\bibnamefont {Xiao}}, \bibinfo {author}
  {\bibfnamefont {Q.}~\bibnamefont {Niu}}, \bibinfo {author} {\bibfnamefont
  {M.}~\bibnamefont {Tsoi}},\ and\ \bibinfo {author} {\bibfnamefont {J.~L.}\
  \bibnamefont {Erskine}},\ }\bibfield  {title} {\bibinfo {title} {Universal
  {Electromotive} {Force} {Induced} by {Domain} {Wall} {Motion}},\ }\href
  {https://doi.org/10.1103/PhysRevLett.102.067201} {\bibfield  {journal}
  {\bibinfo  {journal} {Physical Review Letters}\ }\textbf {\bibinfo {volume}
  {102}},\ \bibinfo {pages} {067201} (\bibinfo {year} {2009})}\BibitemShut
  {NoStop}%
\bibitem [{\citenamefont {Yamane}\ \emph {et~al.}(2011)\citenamefont {Yamane},
  \citenamefont {Sasage}, \citenamefont {An}, \citenamefont {Harii},
  \citenamefont {Ohe}, \citenamefont {Ieda}, \citenamefont {Barnes},
  \citenamefont {Saitoh},\ and\ \citenamefont
  {Maekawa}}]{yamane_continuous_2011}%
  \BibitemOpen
  \bibfield  {author} {\bibinfo {author} {\bibfnamefont {Y.}~\bibnamefont
  {Yamane}}, \bibinfo {author} {\bibfnamefont {K.}~\bibnamefont {Sasage}},
  \bibinfo {author} {\bibfnamefont {T.}~\bibnamefont {An}}, \bibinfo {author}
  {\bibfnamefont {K.}~\bibnamefont {Harii}}, \bibinfo {author} {\bibfnamefont
  {J.}~\bibnamefont {Ohe}}, \bibinfo {author} {\bibfnamefont {J.}~\bibnamefont
  {Ieda}}, \bibinfo {author} {\bibfnamefont {S.~E.}\ \bibnamefont {Barnes}},
  \bibinfo {author} {\bibfnamefont {E.}~\bibnamefont {Saitoh}},\ and\ \bibinfo
  {author} {\bibfnamefont {S.}~\bibnamefont {Maekawa}},\ }\bibfield  {title}
  {\bibinfo {title} {Continuous {Generation} of {Spinmotive} {Force} in a
  {Patterned} {Ferromagnetic} {Film}},\ }\href
  {https://doi.org/10.1103/PhysRevLett.107.236602} {\bibfield  {journal}
  {\bibinfo  {journal} {Physical Review Letters}\ }\textbf {\bibinfo {volume}
  {107}},\ \bibinfo {pages} {236602} (\bibinfo {year} {2011})},\ \bibinfo
  {note} {publisher: American Physical Society}\BibitemShut {NoStop}%
\bibitem [{\citenamefont {Yamane}\ \emph {et~al.}(2016)\citenamefont {Yamane},
  \citenamefont {Ieda},\ and\ \citenamefont {Sinova}}]{yamane_electric_2016}%
  \BibitemOpen
  \bibfield  {author} {\bibinfo {author} {\bibfnamefont {Y.}~\bibnamefont
  {Yamane}}, \bibinfo {author} {\bibfnamefont {J.}~\bibnamefont {Ieda}},\ and\
  \bibinfo {author} {\bibfnamefont {J.}~\bibnamefont {Sinova}},\ }\bibfield
  {title} {\bibinfo {title} {Electric voltage generation by antiferromagnetic
  dynamics},\ }\href {https://doi.org/10.1103/PhysRevB.93.180408} {\bibfield
  {journal} {\bibinfo  {journal} {Physical Review B}\ }\textbf {\bibinfo
  {volume} {93}},\ \bibinfo {pages} {180408} (\bibinfo {year}
  {2016})}\BibitemShut {NoStop}%
\bibitem [{\citenamefont {Chojnacki}\ \emph {et~al.}(2024)\citenamefont
  {Chojnacki}, \citenamefont {Pohle}, \citenamefont {Yan}, \citenamefont
  {Akagi},\ and\ \citenamefont {Shannon}}]{chojnacki_gravitational_2024}%
  \BibitemOpen
  \bibfield  {author} {\bibinfo {author} {\bibfnamefont {L.}~\bibnamefont
  {Chojnacki}}, \bibinfo {author} {\bibfnamefont {R.}~\bibnamefont {Pohle}},
  \bibinfo {author} {\bibfnamefont {H.}~\bibnamefont {Yan}}, \bibinfo {author}
  {\bibfnamefont {Y.}~\bibnamefont {Akagi}},\ and\ \bibinfo {author}
  {\bibfnamefont {N.}~\bibnamefont {Shannon}},\ }\bibfield  {title} {\bibinfo
  {title} {Gravitational wave analogs in spin nematics and cold atoms},\ }\href
  {https://doi.org/10.1103/PhysRevB.109.L220407} {\bibfield  {journal}
  {\bibinfo  {journal} {Physical Review B}\ }\textbf {\bibinfo {volume}
  {109}},\ \bibinfo {pages} {L220407} (\bibinfo {year} {2024})},\ \bibinfo
  {note} {publisher: American Physical Society}\BibitemShut {NoStop}%
\bibitem [{\citenamefont {Wen}\ and\ \citenamefont
  {Zee}(1992)}]{wen_shift_1992}%
  \BibitemOpen
  \bibfield  {author} {\bibinfo {author} {\bibfnamefont {X.~G.}\ \bibnamefont
  {Wen}}\ and\ \bibinfo {author} {\bibfnamefont {A.}~\bibnamefont {Zee}},\
  }\bibfield  {title} {\bibinfo {title} {Shift and spin vector: {New}
  topological quantum numbers for the {Hall} fluids},\ }\href
  {https://doi.org/10.1103/PhysRevLett.69.953} {\bibfield  {journal} {\bibinfo
  {journal} {Physical Review Letters}\ }\textbf {\bibinfo {volume} {69}},\
  \bibinfo {pages} {953} (\bibinfo {year} {1992})}\BibitemShut {NoStop}%
\bibitem [{\citenamefont {Estienne}\ \emph {et~al.}(2023)\citenamefont
  {Estienne}, \citenamefont {Regnault},\ and\ \citenamefont
  {Crépel}}]{estienne_ideal_2023}%
  \BibitemOpen
  \bibfield  {author} {\bibinfo {author} {\bibfnamefont {B.}~\bibnamefont
  {Estienne}}, \bibinfo {author} {\bibfnamefont {N.}~\bibnamefont {Regnault}},\
  and\ \bibinfo {author} {\bibfnamefont {V.}~\bibnamefont {Crépel}},\
  }\bibfield  {title} {\bibinfo {title} {Ideal {Chern} bands as {Landau} levels
  in curved space},\ }\href {https://doi.org/10.1103/PhysRevResearch.5.L032048}
  {\bibfield  {journal} {\bibinfo  {journal} {Physical Review Research}\
  }\textbf {\bibinfo {volume} {5}},\ \bibinfo {pages} {L032048} (\bibinfo
  {year} {2023})}\BibitemShut {NoStop}%
\bibitem [{\citenamefont {Gibbons}\ and\ \citenamefont
  {Werner}(2008)}]{gibbons_applications_2008}%
  \BibitemOpen
  \bibfield  {author} {\bibinfo {author} {\bibfnamefont {G.~W.}\ \bibnamefont
  {Gibbons}}\ and\ \bibinfo {author} {\bibfnamefont {M.~C.}\ \bibnamefont
  {Werner}},\ }\bibfield  {title} {\bibinfo {title} {Applications of the
  {Gauss}–{Bonnet} theorem to gravitational lensing},\ }\href
  {https://doi.org/10.1088/0264-9381/25/23/235009} {\bibfield  {journal}
  {\bibinfo  {journal} {Classical and Quantum Gravity}\ }\textbf {\bibinfo
  {volume} {25}},\ \bibinfo {pages} {235009} (\bibinfo {year}
  {2008})}\BibitemShut {NoStop}%
\end{thebibliography}%

\end{document}